\documentclass[sigconf, nonacm]{acmart}

\usepackage{xpatch}

\makeatletter
\xpatchcmd{\ps@firstpagestyle}{Manuscript submitted to ACM}{}{\typeout{First patch succeeded}}{\typeout{first patch failed}}
\xpatchcmd{\ps@standardpagestyle}{Manuscript submitted to ACM}{}{\typeout{Second patch succeeded}}{\typeout{Second patch failed}}    \@ACM@manuscriptfalse
\makeatother

\renewcommand\footnotetextcopyrightpermission[1]{}
\setcopyright{none}

\AtBeginDocument{%
  \providecommand\BibTeX{{%
    \normalfont B\kern-0.5em{\scshape i\kern-0.25em b}\kern-0.8em\TeX}}}

\usepackage{color}
\usepackage{subcaption}
\usepackage{multirow}
\usepackage{enumitem}
\usepackage[linesnumbered,ruled,vlined]{algorithm2e}
\usepackage{graphicx}

\SetCommentSty{mycommfont}

\SetKwInput{KwInput}{Input}                
\SetKwInput{KwOutput}{Output}              
\SetKw{Continue}{continue}
\SetKw{Break}{break}

\begin{document}

\title{QUEST: An Efficient Query Evaluation Scheme Towards Scan-Intensive Cross-Model Analysis}

\author{Jianfeng Huang}
\email{jfhuang.research@gmail.com}
\affiliation{%
  \institution{Harbin Institute of Technology}
  \city{Harbin}
  \country{China}
}

\author{Dongjing Miao}
\email{miaodongjing@hit.edu.cn}
\affiliation{%
  \institution{Harbin Institute of Technology}
  \city{Harbin}
  \country{China}
}

\author{Xin Liu}
\email{xliu.research@gmail.com}
\affiliation{%
  \institution{Harbin Institute of Technology}
  \city{Harbin}
  \country{China}
}

\renewcommand{\shortauthors}{Huang and Miao, et al.}
\newenvironment{claim}[1]{\par\noindent\textbf{Claim}:\space#1}{}



\begin{abstract} Modern data-driven applications require that databases support fast cross-model analytical queries. Achieving fast analytical queries in a database system is challenging since they are usually scan-intensive (\textit{i.e.}, they need to intensively scan over a large number of records) which results in huge I/O and memory costs. And it becomes tougher when the analytical queries are cross-model. It is hard to accelerate cross-model analytical queries in existing databases due to the lack of appropriate storage layout and efficient query processing techniques. In this paper, we present {QUEST} (\underline{QU}ery \underline{E}valuation \underline{S}cheme \underline{T}owards scan-intensive cross-model analysis) to push scan-intensive queries down to unified columnar storage layout and seamlessly deliver payloads across different data models. QUEST employs a columnar data layout to unify the representation of multi-model data. Then, a novel index structure named \textit{Skip-Tree} is developed for QUEST to enable the query evaluation more efficient. With the helps of two pair-wise bitset-based operations coupled with \textit{Skip-Tree}, the scan of most irrelevant instances can be pruned so as to avoid the giant intermediate result, thus reducing query response latency and saving the computational resources significantly when evaluating scan-intensive cross-model analysis. The proposed methods are implemented on an open-source platform. Through comprehensive theoretical analysis and extensive experiments, we demonstrate that QUEST improves the performance by $3.7\times -\ 178.2\times$ compared to state-of-the-art multi-model databases when evaluating scan-intensive cross-model analytical queries.  
\end{abstract}






\maketitle

\section{Introduction}
In addition to the conventional relational data model, document-based and graph-based data models have become indispensable types of information in modern data-driven applications.
Joint analysis across different model type of datasets has been a very important function for modern data-driven applications \cite{lu2019multi}.
By conducting joint analysis on diverse datasets, valuable insights can be extracted from the interplay among different models, enabling more comprehensive optimization of business decisions and unlocking the full potential of big data \cite{zhang2019unibench}.
As discussed in \cite{zhang2019unibench,bimonte2023logical}, cross-model analysis performs typical scan-based analytical operations, such as filtering, aggregating and grouping, on each individual type of data respectively and integrates the information by joining across different data model types.
Different from developing techniques for the analytical query processing in a single data model, such as for relational model \cite{abadi2006integrating,abadi2009column,abadi2006materialization}, for document-based model \cite{melnik2010dremel,afrati2014storing,wen2019cores,wang2017exploiting} and for graph-based data\cite{angles2020ldbc,gupta2021columnar,jin2021making,francis2018cypher},
to achieve fast cross-model analysis, it is necessary to consider how to process analytical queries efficiently on each type of data at the same time and how to compute join efficiently across different model types \cite{chen2018worst,chen2022cross}.

\begin{figure*}[h]
  \centering
  \includegraphics[width=1\linewidth,height = 6 cm]{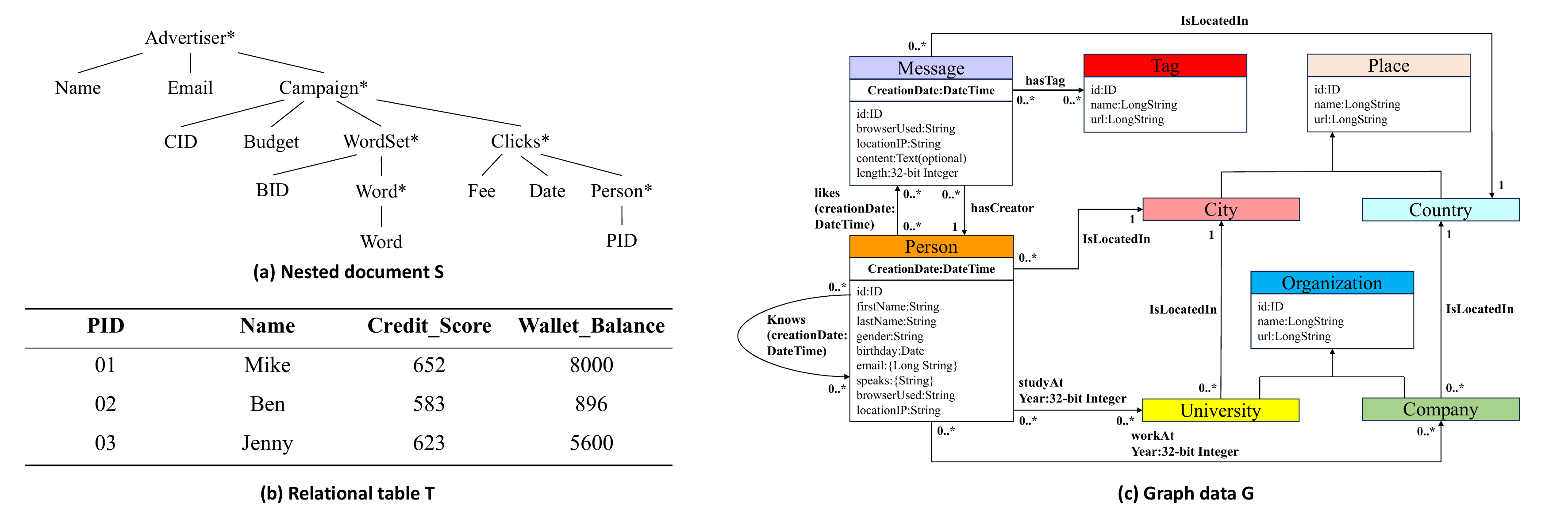}
  \caption{Motivating example of multi-model data}
  \label{example}
\end{figure*}

Consider the following scenario, a popular social media platform with a substantial user base collaborates with numerous businesses for advertising purposes.
A natural way is to model people’s social network data as a property graph \cite{angles2020ldbc},
model the personal information (\textit{e.g.}, credit score and wallet balances on the platform) as a table,
and employ a nested tree structure to effortlessly model the advertisers’ campaigns which are usually recorded in JSON files \cite{melnik2010dremel,afrati2014storing}.
In this way, the schema can be built as shown in Figure~\ref{example}\footnote {Note that, $A^*$ denotes the attribute $A$ can be repeated zero or more times. where we utilize the symbolic representation for tree data in DREMEL \cite{afrati2014storing}. }. 


In these multi-model data the platform may be demanded to select the candidate users for promotional purposes.
To this end, a query $Q$ is described here to select ``all the users who is counted into the \textit{clicks} of the \textit{campaign} that comprises the \textit{wordset} containing \textit{word} $w_1$ AND whose \textit{credit\_score} is higher than $s_1$ AND whose friend (\textit{i.e.,} the person they \textit{know}) \textit{liked} the \textit{message} with \textit{tag} $t_1$''.
Note that, query $Q$ is a scan-intensive cross-model analytical query because, to evaluate this query, a large number of records should be scanned and different types of data should be joined.
Specifically, $Q$ includes three filters ``$\textit{\text{credit\_score}}>s_1$'' on table $R$, ``$\textit{\text{tag}}=t_1$'' on graph $G$ and ``$\textit{\text{Advertiser.Campaign.WordSet.Word}}=w_1$'' on document $S$,
meanwhile, $Q$ includes a pattern match performing on \textit{know} and \textit{like} edges in graph and two cross-model joins through Person's ID to integrate information.

Such scan-intensive cross-model analytical query $Q$ can be evaluated using various approaches depending on the chosen storage strategy. Either the data is stored in different engines corresponding to the four data models, or the three types of data are transformed into a unique format, and stored in a fully integrated single-engine backend.
For example, one can transform $S$, $R$ and $G$ into a unique format, say table, and management them in a relational database.
However, in this way, duplicate data is required across multiple tables to represent the nested hierarchy of $S$ and the adjacent relations between vertexes in $G$. Furthermore, when evaluating query $Q$, it requires joining multiple tables or using recursive queries to navigate through nested records and traverse in graph $G$ which may result in sever response latency and memory crash. Likewise, the other existing schemes fall short to efficiently evaluate analytical queries on each type of data at the same time due to the heterogeneity of multi-model data.

In fact, it's challenging for existing databases to efficiently evaluate the scan-intensive cross-model queries since the inappropriate storage layout and inefficient query processing techniques may results in huge I/O and memory costs. The main bottlenecks can be drawn out as follows:

\textbf{Storage layout.}
The mainstream storage layout for multi-model data could be generally divided into two categories \cite{lu2019multi}, the multi-engine federated databases based on middleware and the single-engine multi-model databases based on a integrated backend. When processing the scan-intensive cross-model, the existing modern federated databases have to scan over a large number of records stored at distinct places and join them, which may result in extra data copy, migration and integration costs \cite{lu2019multi,kiehn2022polyglot,duggan2015bigdawg}. Although the existing single-engine multi-model databases based on unique model achieve unified storage and management of multi-model data, they still face the ``One size doesn't fit all'' dilemma, and the inappropriate data layout will result in loss of performance and flexibility for specific cross-model analytical queries \cite{zhang2019unibench, azevedo2020modern,van2023comparative}. Hence, the existing multi-model data storage methods fall short to efficiently support scanning-intensive cross-model analysis.

\textbf{Query evaluation.}
When evaluating cross-model analytical queries,  the existing multi-engine federated databases decompose the query into multiple sub-queries, assign them to individual storage engines for separate processing through multiple query interface. Then collect the dispersedly results and integrate the final output through middleware assembly \cite{duggan2015bigdawg,azevedo2020modern}. The process results in expensive query language coding, compiling and translating cost due to the heterogeneity of different engines. It will greatly limit the flexibility of query execution and retards the efficiency of evaluating cross-model analytical queries. Although the existing single-engine multi-model databases are able to evaluate cross-model analytical queries through unified interface,  they may still scan a large number of irrelevant instances due to the lack of customized evaluation scheme which results in huge extra I/O and memory costs \cite{zhang2019unibench,van2023comparative}. Meanwhile, both types of existing databases may generate giant intermediate results and huge memory cost due to the lack of efficient cross-model join algorithm. Hence, the existing query processing techniques fall short to efficiently evaluate the scanning-intensive cross-model queries.

\textbf{Query optimization.}
The multi-engine federated databases' query execution is limited when dealing with cross-model analysis, and fall short to perform fine-grained optimization 
 \cite{azevedo2020modern,duggan2015bigdawg}. It's also challenging for single-engine multi-model databases to establish a comprehensive costs model and customize optimization method towards cross-model analysis, since it's non trivial to present a comprehensive and appropriate logical model for multi-model data \cite{holubova2021multi}. Thus, the existing query optimization methods fall short to achieve the optimal query execution when dealing with specific cross-model analysis which results in severe response latency and even process crash \cite{zhang2019unibench,van2023comparative}.

To break through the bottlenecks mentioned above,
we present our solution QUEST which is an efficient query evaluation scheme towards scan-intensive cross-model analysis. Specifically, QUEST aims to tackle the challenges of joint analysis on relational table, document-based data that can be modeled as nested tree-structure (\textit{e.g.}, JSON files and XML files) and graph-based data that can be modeled as property graph. The key idea behind QUEST is to leverage columnar storage layout and advanced column-oriented techniques to develop customized evaluation scheme for scan-intensive cross-model queries. The main contributions of this paper are as follows:

\begin{itemize}[leftmargin=*]

\item{We employ columnar data layout to unify the representation of multi-model data.
Specifically, we first unify the logical model of the relational,
nested document-based data and property graph-based data based on the extended recursive definition of nested tree-structured model.
And develop a lossless representation of record structure in a columnar format.
\textit{Counter} and \textit{Indicator} arrays stored in columns are developed to maintain the mapping information between adjacent layers in nested model. A novel index structure \textit{Skip-Tree} is developed to preserve the pre-computed mapping information across nested layers to enable the query evaluation more efficient.}

\item{We present a novel column-oriented skipping scheme based on \textit{Skip-Tree} structure and bitset-based query storage pushdown strategy, generalized by a two pair-wise operations, $\textit{SkipUp}$ and $\textit{SkipDown}$. It can significantly reduce I/O and CPU cost when evaluating scan-intensive cross-model queries by pruning the scan of most irrelevant instances. We also introduce a way to seamlessly deliver query payloads across different data models to avoid the giant intermediate result caused by cross-model joins.}

\item{We delve into the query evaluation costs of QUEST and establish a comprehensive cost model that encompasses both I/O and CPU cost. As for the query optimization, we establish correctness constraints for predicates evaluating order in nested model, which enables us to analyse the computational complexity and explore efficient algorithms for solving the optimal predicates evaluating order.}

\item{We scrutinize the characteristics of data access patterns in scan-intensive cross-model queries and analyse the primary factors that decrease database's efficiency when evaluating such queries. These findings further evolve into our refinement of choke points for cross-model analytical workloads, motivating us to generate new multi-model dataset and corresponding micro benchmark.} 
\end{itemize}

The rest of this paper is organized as follows:
In section 2, We discuss the unified nested modeling and columnar storage layout for multi-model data.
Section 3 and 4 present the query evaluating and theoretical optimizing techniques in detail, respectively.
Extensive experiments are conducted in Section 5.
The last two sections are related works and conclusions of this paper. 

\section{COLUMNAR DATA LAYOUT}
\subsection{Nested Modeling of Multi-Model Data}
QUEST tends to employ columnar data layout to unify the representation of relational, nested document-based data and property graph-based data so as to streamline data retrieval and facilitate cross-model query evaluation. 
Specifically, relational tables can be modeled as a set of columns through vertical partition \cite{abadi2006integrating,abadi2009column,abadi2006materialization}. 
And the nested document-based data can be modeled as a set of columns which are organized as the same tree hierarchy as its data schema \cite{melnik2010dremel,wang2017exploiting,wen2019cores}. 
Although the property graph-based data can also be modeled as a set of columns in a trivial way, 
it will cause huge storage redundancy and memory crash due to the intricate mapping and circular relationships among vertexes (\textit{e.g.}, the one-to-many, many-to-many and many-to-one mappings)  \cite{gupta2021columnar,besta2019demystifying}. 
Thus, an efficient pruning scheme is needed to lossless represent its structure in a columnar format. 
QUEST addresses this problem by extending the recursive definition for nested tree-structured data model \cite{wang2017exploiting} to enable an efficient pruning on property-graph based data so as to unify the both the logical and physical representation of multi-model data based on columnar data layout.

\begin{align*}
\begin{split}
&T_{value}=T_{record}\:|\:T_{array}\:|\:T_{primitive}\:|\:T_{indicator},\\ 
&T_{record}=\{key_1:T_{value_1},\dots,key_n:T_{value_n}\},\\
&T_{array}=[T_{value},\dots,T_{value}],\\
& 
T_{indicator}=\textit{\text{pointer}}\ to\ T_{value}, \\
&T_{primitive}=string\:|\:number\:|\:boolean\:|\:null,\\
&key_i|i\in [1,n]=string,\\
&T_{root}=T_{record},\\
&T_{leaf}=T_{primitive}\:|\:T_{indicator}.
\end{split}
\end{align*}

The extended recursive definition of nested tree-structured data is shown above. In this definition, a $record$ has a unique $root$ which is composed of a collection of fields, each distinguished by a field name as $key$ and a pre-defined field type as $T_{value}$. A field in a $record$ can be defined as one of the $primitive$ types, such as $string, number, boolean$ or $null$. Alternatively, it can be recursively defined as either another $record$ or $array$ type. Especially if the parent is of array type, there exists a list of child instances with the same type embedded in a common parent instance, denoted as orderly one-to-many relationship. In this case, for each instance of a parent node, there may exist more than one instance in its child node. In addition, we introduce a novel field type: $T_{indicator}$ which enables the nested model to depict the intricate many-to-one relationships within the property graph. The $indicator$ type of node serves as a pointer to the instances of specific $T_{value}$. Especially if the parent is $indicator$ type, multiple parent instances with the same type may share a common child instance. The field type of leaf node on nested tree can only be $indicator$ or $primitive$.
In all of the field types formalized above, only the units of $T_{primitive}$, $T_{indicator}$ and $T_{array}$ have instances in storage. Specially, the instance of all types of fields are stored in columnar format and organized as the same tree hierarchy as the extended recursive definition in QUEST.  

\begin{figure}[h]
\centering
  
\includegraphics[height = 4cm]{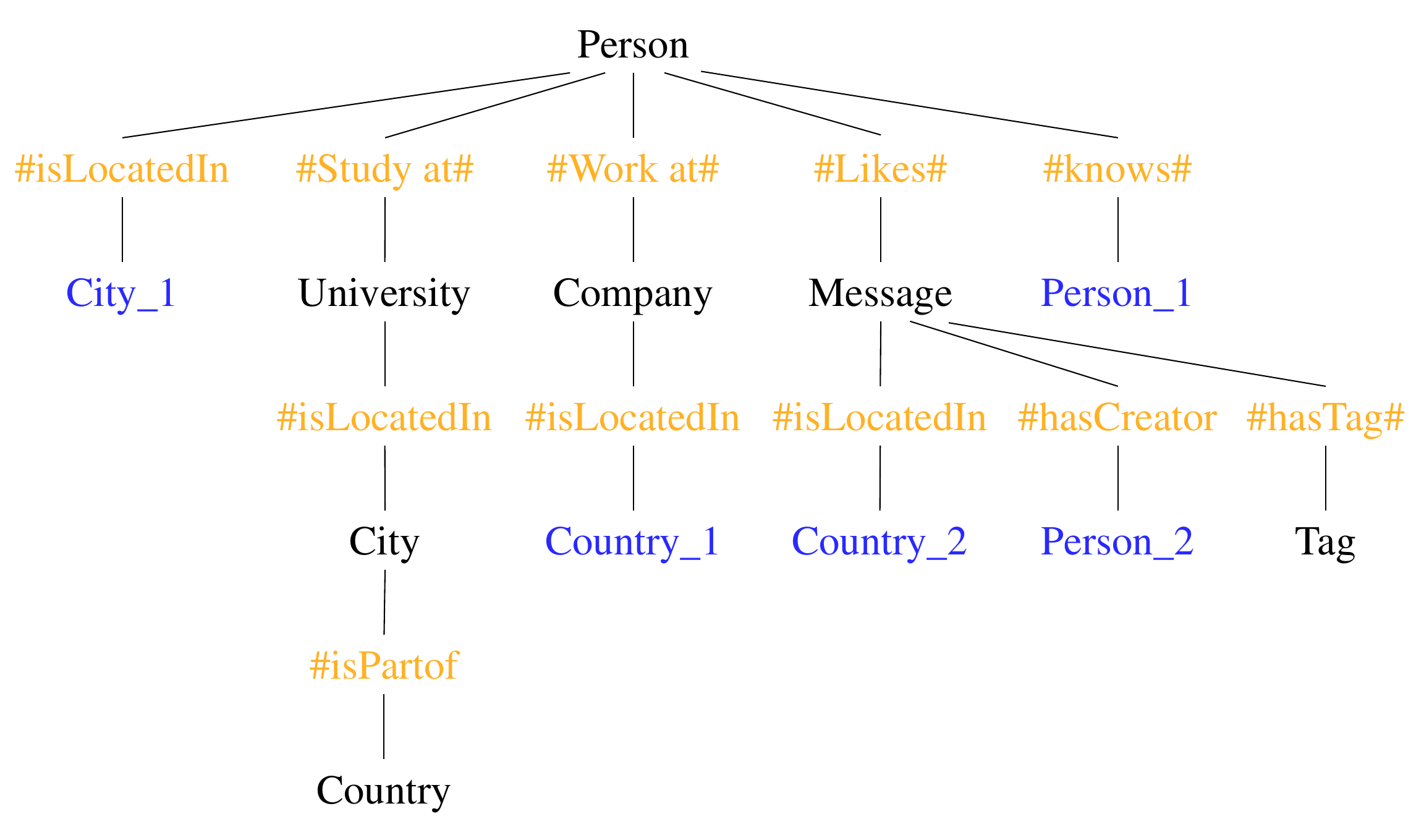} 
\label{figure:subim1}
\caption{The nested schema of Figure \ref{example} (c).}
\label{schema}
\end{figure}

We demonstrate the immense potential of our extended definition of nested structures based on the example of multi-model data in Figure \ref{example}. The schema of nested document shown in \ref{example} (a) could be depicted based on the basic recursive definition of tree-structured data. As for the property graph shown in Figure \ref{example} (c), starting from the \textit{Person} node, one could obtain a nested tree structure presented in Figure \ref{schema} by applying the classic traversal algorithm on graph like depth-first searching. During the searching process, $T_{indicator}$ are utilized to mark the previously traversed vertexes as leaf nodes, effectively pruning that branch of the tree. For the sake of simplicity, we omit all fields of primitive type in Figure \ref{schema}. The orange nodes represent the edges in the property graph where the many-to-many mapping is denoted ``$\#E\#$'', the many-to-one mapping is denoted ``$\#E$'' and the one-to-many mapping  is denoted``$E\#$''.

The $indicator$ type of fields not only can efficiently prune the branch, but also play a pivotal role to depict the many-to-many mapping relationship within the property graph. Specifically, QUEST accords equal importance as $T_{value}$ to the property of vertexes and edges in nested tree structure, thereby separating the intricate many-to-many relationship between connected vertexes into the more manageable one-to-many vertex-to-edge relationship and the many-to-one edge-to-vertex relationship. Thus, the many-to-many mapping information could be efficiently maintained by a $T_{array}$ in conjunction with a $T_{indicator}$. Consequently, vertexes in the graph can be regarded as $T_{record}\:|\:T_{indicator}$ node or a combination of the two in the nested tree structure which depends on its in-degree and out-degree. And the edges can only be regarded as $T_{array}$ since the vertex-to-edge relation can only possible be one-to-many or one-to-one. Overall, the extended recursive definition of nested tree based on $indicator$ type of fields enables QUEST to employ columnar layout to unify the representation of relational, nested document-based data and property graph-based data.

\subsection{Unified Tree Metadata Management}
As previously mentioned, only $T_{\textit{\text{primitive}}}\:|\:T_{\textit{\text{array}}}\:|\:T_{\textit{\text{indicator}}}$ types of fields in nested tree structure have instances in storage. Among these, instances of $T_{primitive}$ could be naturally stored in columnar format. Therefore, the crux to efficiently support scan-intensive cross-model analysis lies in how to maintain the mapping information between nested layers contained by $T_{\textit{\text{array}}}$ and $T_{\textit{\text{indicator}}}$. In contrast to the encoding of global structural information in tree structures \cite{afrati2014storing,alkowaileet2022columnar}, we opt to disentangle the global structure and concentrate solely on the mapping information between nested layers. The ``parent-child'' relationship on nested tree structure is preserved as metadata. This approach enables QUEST to query nodes in arbitrary nested depth without reconstructing the overall nested structure, thereby to fully leverage columnar layout's advantages in data retrieval. 

\begin{figure}[h]
\centering
\begin{subfigure}{0.5\textwidth}
\includegraphics[width= \linewidth,height =3cm]{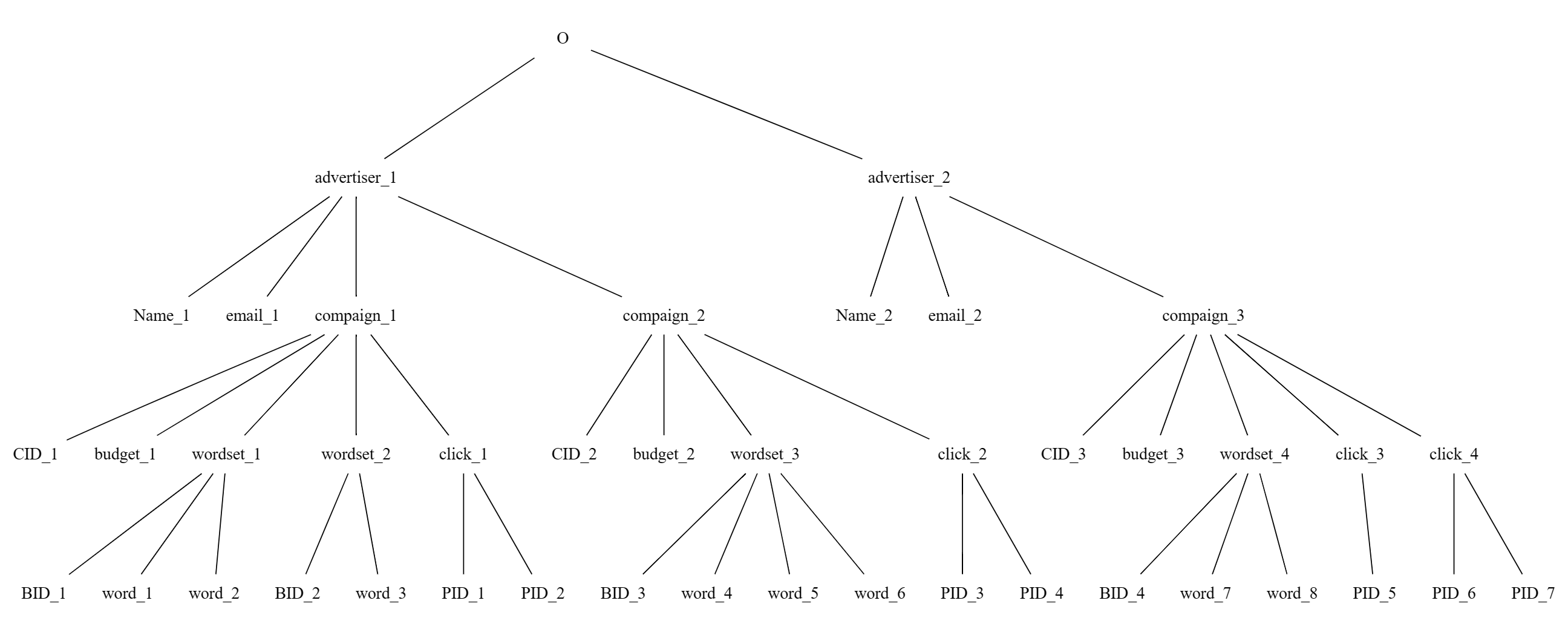} 
\caption{Example Instance of document $S$ (Schema shown in Figure \ref{example} (a))}
\label{figure:subim1}
\end{subfigure}
\begin{subfigure}{0.5\textwidth}
\centering
\includegraphics[width= 0.6\linewidth,height= 3cm]{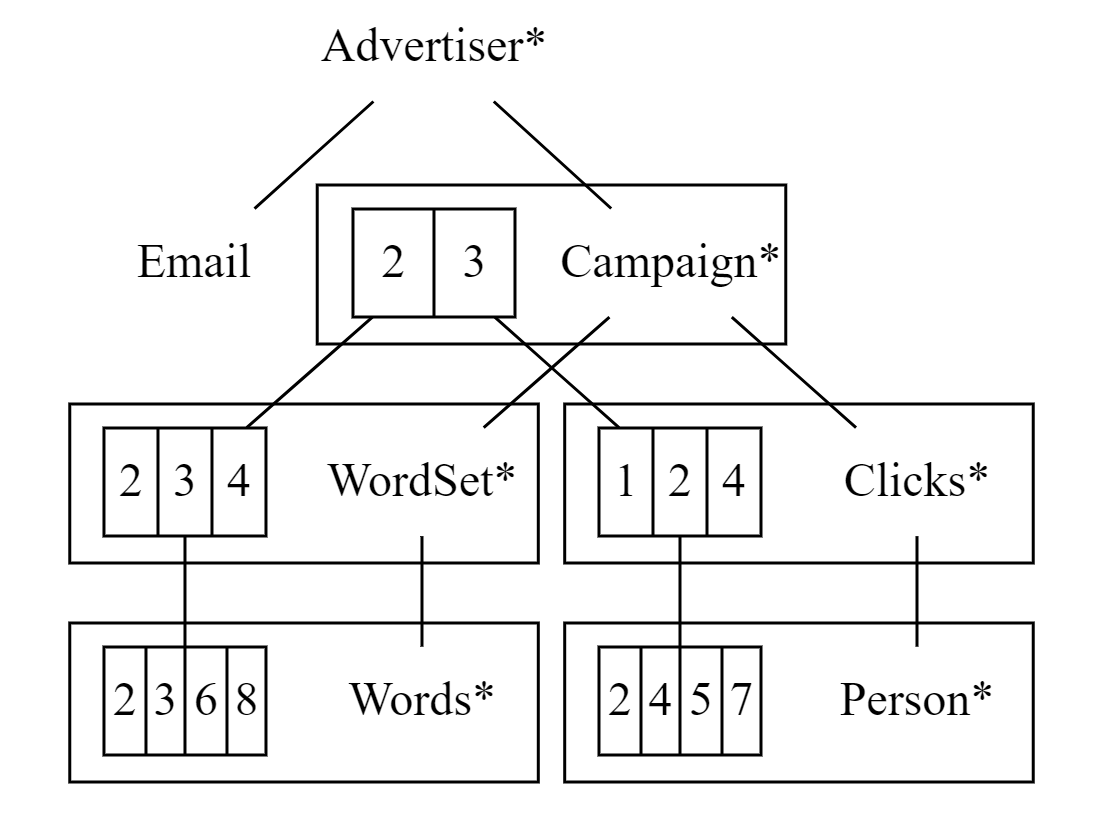}
\caption{The tree metadata based on the \textit{Counter} of (a)}
\label{fig:subim2}
\end{subfigure}
\caption{Illustration for tree metadata management}
\label{metadata1}
\end{figure}

Specifically, QUEST stores the instances of $T_{\textit{\text{array}}}$ and $T_{\textit{\text{indicator}}}$ in columnar format as well. For $T_{\textit{\text{array}}}$, similar with \cite{wen2019cores}, we add a corresponding index \textit{Counter} to the columnar storage to represent the nested relationships between layers, \textit{i.e.}, an array to record the position in the column of the last child instance that belongs to its parent. These arrays are stored in columnar format and  organized as the same tree hierarchy as the nested data schema based on the extended recursive definition. For instance, Figure \ref{metadata1} (a) shows the specific instances of the nested document illustrated in Figure \ref{example} (a), there are two \textit{Advertiser} instances $a_1,a_2$ and three \textit{Campaign} instances $c_1,c_2,c_3$, stored sequentially in columnar storage, the \textit{Counter}$\textit{[2,\ 3]}$ represents that $c_1,c_2$ was organized by $a_1$ and $c_3$ was organized by $a_2$.  The \textit{Counter} array with the same size as the parent's cardinality records the offset range of child instance in columns which enables swift location in the columnar storage when traversing from nonleaf instances to its descendants.

\begin{figure}[h]
\centering
  
\begin{subfigure}{0.5\textwidth}
\centering

\includegraphics[width= 0.85\linewidth,height =4.5cm]{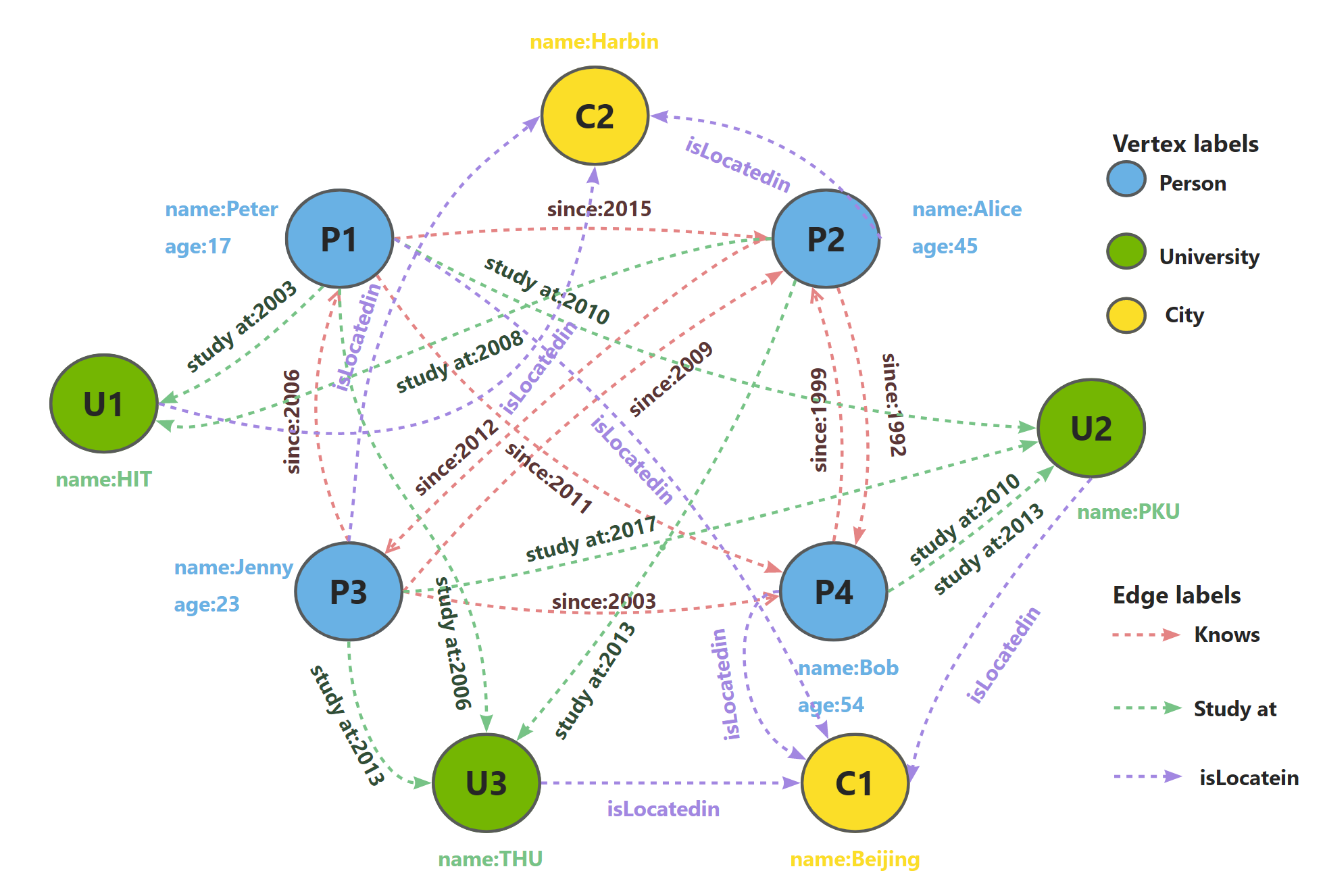} 
\caption{Example instance of partial graphs in social networks}
\label{figure:subim1}
\end{subfigure}

\begin{subfigure}{0.5\textwidth}
\centering

\includegraphics[height= 3.5cm]{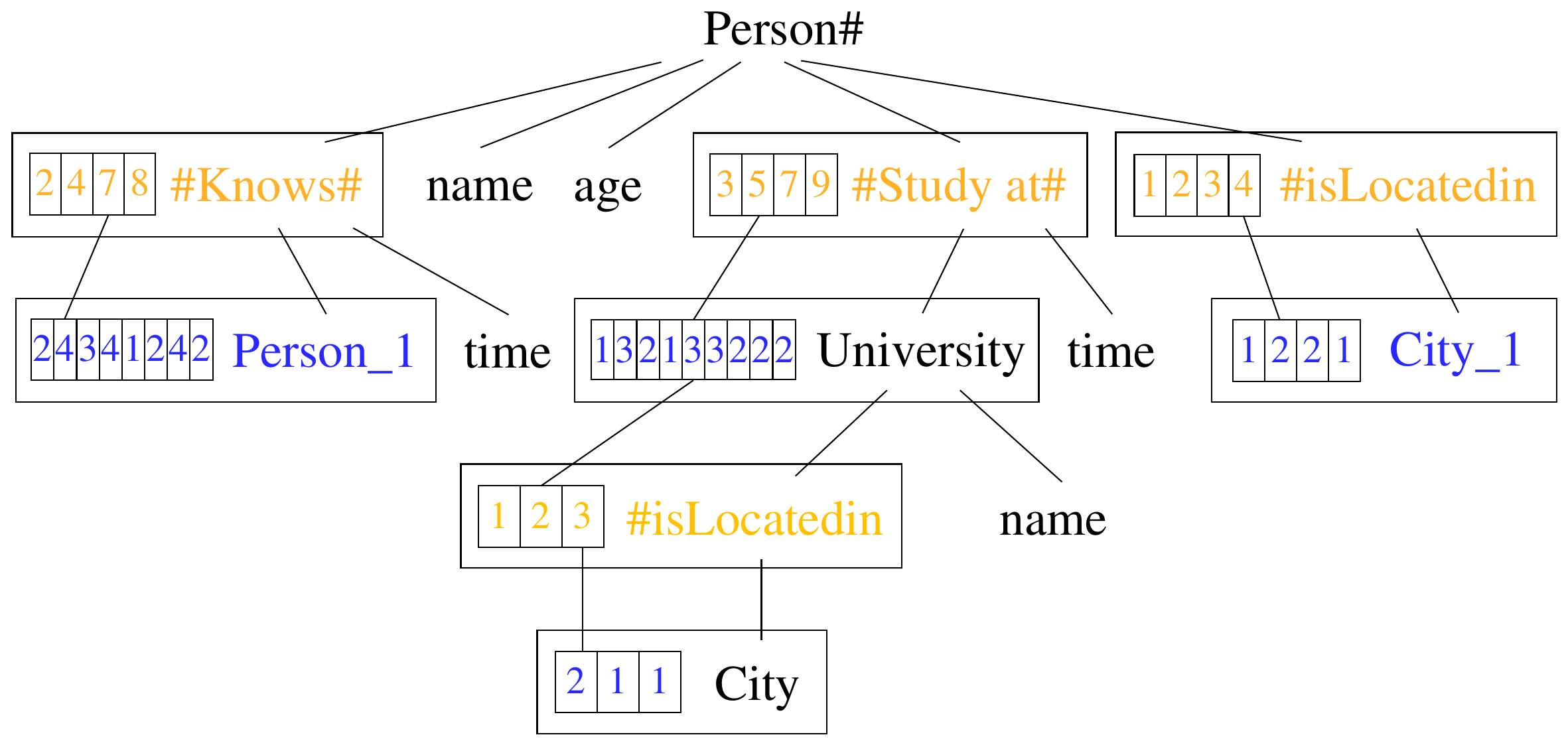}
\caption{Nested tree metadata of (a)}
\label{fig:subim2}
\end{subfigure}

\caption{Illustration for tree metadata management}
\label{metadata2}
\end{figure}

As for property graph-based data instance shown in Figure \ref{metadata2}, the one-to-many mapping information contained by the $array$ type of field is still preserved within the \textit{Counter} arrays. Furthermore, QUEST adopts an \textit{Indicator} array to store the instances of  $indicator$ type of field in columns as well. The \textit{Indicator} array with the same size of the child's cardinality records the precise offset of each child instance in column. The combination of \textit{Counter} and \textit{Indicator} array can be essentially viewed as a CSR form of adjacency matrix on graphs. And the introduction of \textit{Counter} in CORES \cite{wen2019cores} for nested document data can be viewed as a natural simplification of CSR structure to facilitate data retrieval in one-to-many mapping nested relationship. QUEST employs nested columns to provide a uniformed logical frame based on the extended recursive definition of nested tree-structured model. We next demonstrate that integrating the multi-model data into unified tree logical model based on columnar layout enables QUEST to develop an efficient skipping scheme towards scan-intensive cross-model analysis.

\subsection{Precomputed \textit{Skip-Tree} Index Structure}

\subsubsection{Precomputation on nested documents}
We acknowledge that when processing conjunctive queries in tree-structured data, it is necessary to locate the lowest common ancestor (LCA) of two nodes to transmit intermediate results \cite{wen2019cores, afrati2014storing}. However, transmitting the intermediate results layer-by-layer in nested structure is time-consuming, as each step incurs additional I/O and memory costs. Consequently, when processing queries on data with deeper nesting levels, analytical query latency will be significantly increased. Nevertheless, if we precompute some  selective ancestor-descendant mapping information and organize it in the form that is similar to the preprocessing process of binary lifting algorithm, we can mitigate these issues. Thus QUEST develops the \textit{Skip-Counter} array to maintain the precomputed  mapping information and organize them as a novel data structure called \textit{Skip-Tree} which is inspired by the idea of \textit{Skip-List}.

\begin{figure}[h]
\centering
\includegraphics[width=0.8\linewidth]{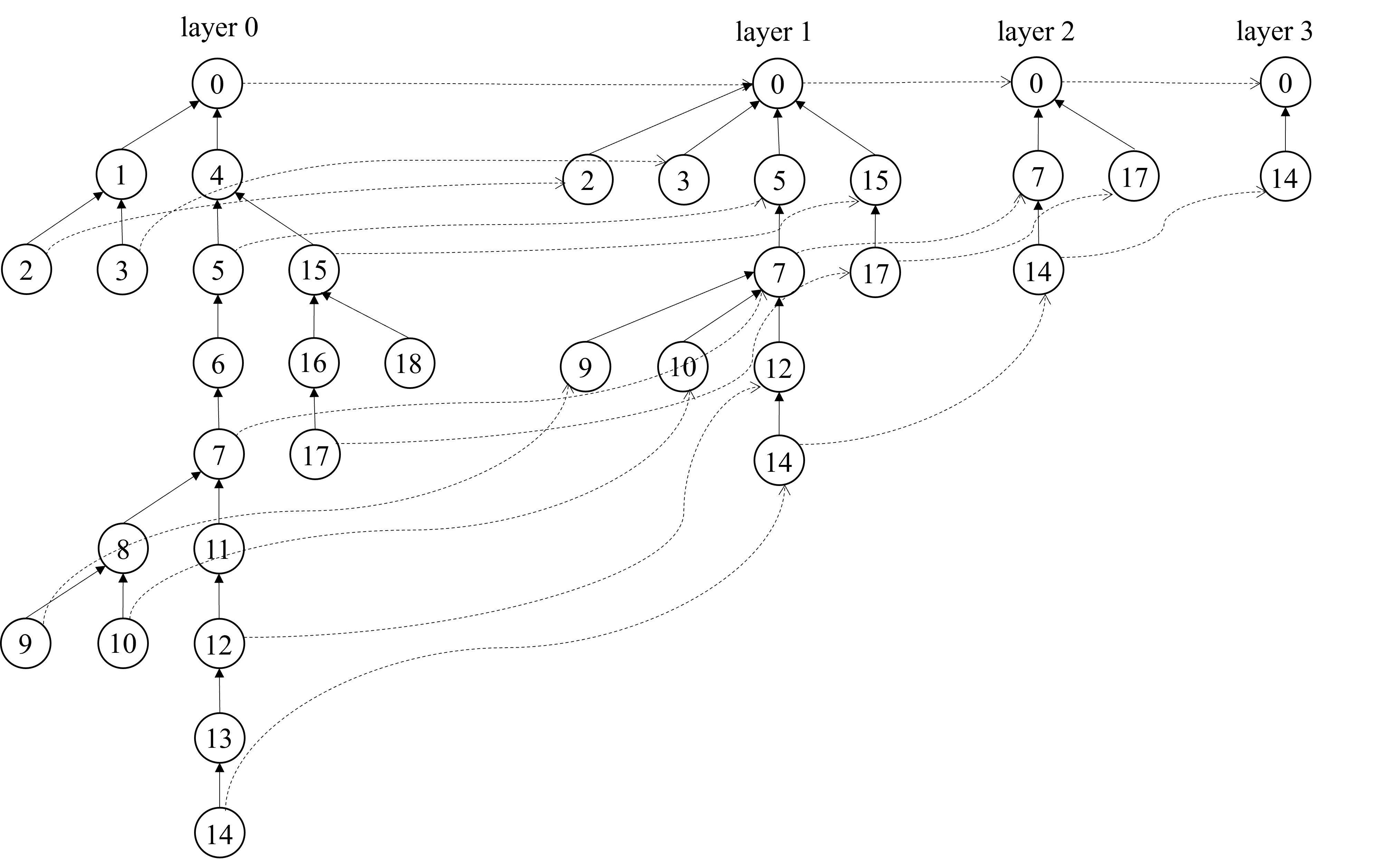}
\caption{Illustration for \textit{Skip-Tree} structure}
\label{skiptree}
\end{figure}

A \textit{Skip-Tree} is a tree-based lookup structure that leverages the concept of \textit{Skip-List}. The \textit{Skip-Tree} consists of multiple layers, with a bottom layer containing all nodes in the tree where each node maintains a pointer to its parent node. The probability of each node to be lifted up to the next level is set to $\frac{1}{2}$ in each level of \textit{Skip-Tree}. We denote the height of a node in the \textit{Skip-Tree} as the highest level that includes the node (with the bottom of the \textit{Skip-Tree} being denoted as height 0). For every node with the height greater than 0, we maintain a \textit{Skip-Ancestor} list to preserve their nearest ancestor at each lower level than its height. Clearly, the desired space overhead of a \textit{Skip-Tree} constructed on a tree containing $n$ nodes with depth of $d$ is $O(n)$, and the desired time complexity of seeking the LCA can be proved to be $O(\log d)$. However, when it is applied to  realistic use with \textit{Skip-Counter}, the actual efficiency promotion is related to the scale of data instances. The \textit{Skip-Tree} depicted in Figure \ref{skiptree} is constructed from a tree comprising of 18 nodes, and only the nodes with odd depths are lifted up to higher layers. This fixed construction for static schema structure ensures stable time overhead when searching for specific ancestors in the \textit{Skip-Tree}, making it more suitable for evaluating scan-intensive analytical queries. Therefore, the present implementation of QUEST also employs the same static schema structure.

In addition, during the process of initializing the \textit{Skip-Tree} structure, QUEST not only records each node's \textit{Skip-Ancestor} list of the first ancestor at different levels lower than its height, but also pre-compute the potential one-to-many ancestor-descendant mapping information and preserved as \textit{Skip-Counter}. For example, in the nested model shown in Figure \ref{metadata1} (b), the \textit{Counter} of \textit{Clicks} is \textit{[1, 2, 4]} and the \textit{Counter} of \textit{Person} is \textit{[2, 4, 5, 7]}, thus the \textit{Skip-Counter} \textit{[2, 4, 7]} is expected to be calculated when the ancestor \textit{Campaign} 
 is concatenated to \textit{Person}'s \textit{Skip-Ancestor} list, indicating that the three \textit{Campaign} instances corresponds to the 1st-2nd, 3rd-4th, and 5th-7th \textit{Person} instances respectively. The specific initial process is shown in algorithm \ref{alg:algorithm1}.

\begin{figure}[h]
\centering
\includegraphics[width= 0.55\linewidth, height = 3cm]{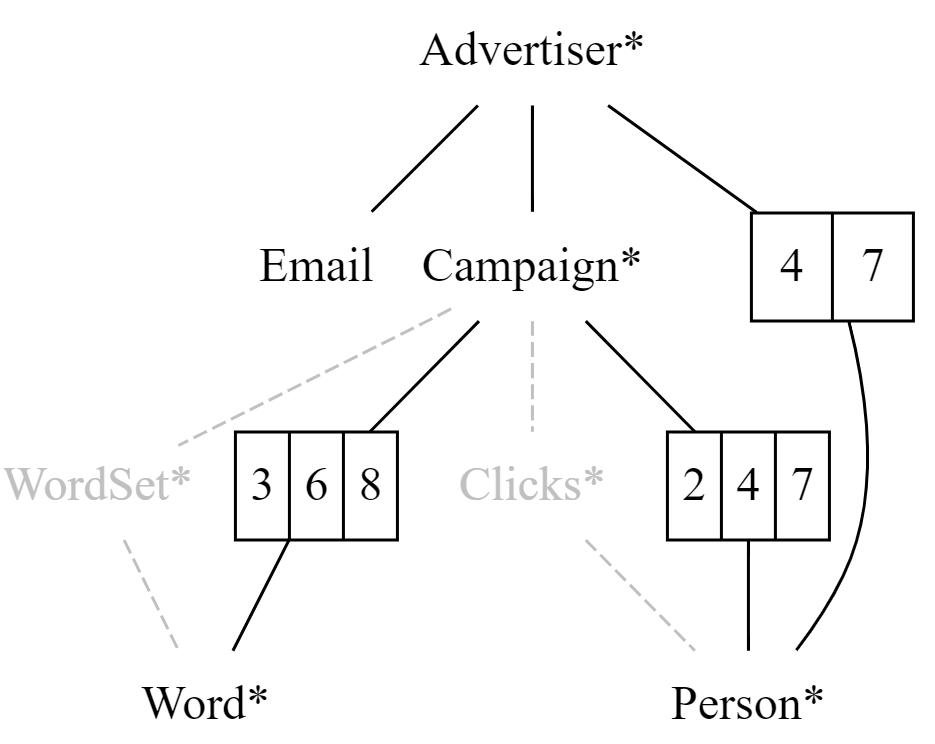}
  \caption{Illustration for \textit{Skip-Counter} structure}
  \label{skipcounter}
\end{figure}

\begin{algorithm}[]
\DontPrintSemicolon

  \KwInput{$T:$ Tree schema; $S:$ Nodes in $T$; $D:$ the max depth of $T$;
  $H:$ the max height of \textit{Skip-Tree} index $= \lceil \log_2 D \rceil$.}
  
  \KwOutput{$ST:$ \textit{Skip-Tree} Index structure.}
  \SetKwFunction{FSetHeight}{SetHeight}
  \SetKwProg{Fn}{Function}{:}{}
   \Fn{\FSetHeight{$T$}}{
 \textit{$s.depth$} = the node's depth in $T$; \textit{$s.height$} =$0$; \;
\For{$s$ in $T$}
{\For{$i$ in 1\dots $H$}
    {
     \If\textit{{$s.depth$} mod $\ 2^i==0$}{\textit{$s.height$} $++$;}
    }
  }
  }

  \SetKwFunction{FSkipTree}{SkipTree}
  \SetKwProg{Fn}{Function}{:}{}
   \Fn{\FSkipTree{$T$}}{
   
\For{$s$ in $T$}
{ \textit{$s.SkipAncestor$} $=[]$; \textit{$s.SkipCounter$}  $=[];$\;
\textit{$Counter$} $=$ the \textit{Counter} array of $s$;\;
  \For{$j$ in 0\dots \textit{$s.height$}}
    { $s_j=$ the nearest ancestor of $s$ whose height $\geq j$; \;
 \textit{$s.SkipAncestor.append(s_j)$};\;
     \textit{$NewCounter$} = $ \emph{CounterUnion} (s_j, s)$; \;
     \tcp{Calculate the \textit{Skip-Counter}}
    \textit{$s.SkipCounter.append(NewCounter)$};}
  }
  }
  
\caption{Precomputation on nested document}
\label{alg:algorithm1}
\end{algorithm}

\subsubsection{Precomputation on graphs}
However, the one-dimensional precomputation falls short in capturing the intricate many-to-many mapping between nodes in graph. According to our fixed \textit{Skip-Tree} structure design, in the property graph-based data, only the vertexes will be lifted up to higher layers which means that the height of all edges is set to 0. Therefore, the precomputation process solely focus on vertexes in the original graph. As previously mentioned, the \textit{Counter} embedded with \textit{Indicator} is equivalent to the CSR format of adjacency matrices between vertexes. As such, the precomputing process involves restoring the CSR structure based on \textit{Counter} and \textit{Indicator} to adjacency matrix, followed by computing the multi-hop adjacency matrix through two-dimensional operations akin to matrix multiplication. We omit the details since it's basic and well-known. In addition, the multi-hop matrix is further condensed into new CSR format (\textit{i.e.,} a \textit{Skip-Counter} with a \textit{Skip-Indicator}). Note that this process is also subject to our uniform nested modeling and \textit{Skip-Tree} pre-computation strategy, that is, in the nested tree structure expanded by the graph, the precomputation process is completed when the \textit{Skip-Tree} structure is initialized.

\section{SCAN-INTENSIVE CROSS-MODEL ANALYTICAL QUERY EVALUATION}
\subsection{Bitset-based Filter Storage Pushdown}
This section introduces a predicates pushdown strategy based on bitsets, which enables QUEST to support conjunctive filters on arbitrary paths within a nested schema. Multiple filtering predicates are considered with conditions related to any set of fields. The bitset-based storage pushdown scheme enables the upcoming operator only to deal with the valid units that have been reserved so far based on previous filters. We demonstrate how to navigate from the current node to a desired reachable node in order to deliver decreasing payload to the next operator.

The main idea is to iteratively deliver a lightweight bitset from the current operator to an upcoming one. Thus, the latter can take over the valid payloads from the former. In practice, the bitset can be realized with a bit vector (\textit{e.g.}, BitSet in Java). With ``1'' for hitting and ``0'' otherwise, each bit denotes whether a corresponding unit has been chosen so far. The process can be abstracted by two basic compositions, \textit{i.e.,} $\textit{RollUp}$ and $\textit{DrillDown}$ 
\cite{wen2019cores}. Given a one-to-many relationship, $\textit{RollUp}$ delivers a bitset from a child node to its array-typed parent. Reversely, if a bitset need to be delivered from the parent to one of its children, $\textit{DrillDown}$ will hit a range of units in the child’s bitset based on the status in their parent's bitset. 

The basic bitset delivery operations in property graph-based data is similar. When transmitting the bitset through one-to-many vertex-to-edge relationships which are uniformly stored in \textit{Counter} arrays, QUEST directly utilize $\textit{RollUp}$ and $\textit{DrillDown}$ operations. When it comes to transmit the bitset through  many-to-one edge-to-vertex relationships which are uniformly stored in \textit{Indicator} arrays, the every bit in parent's bitset will be generated by retrieving the bit in child's bitset pointed by the corresponding unit of \textit{Indicator}. Combining the above two delivery operations, the bitset can be smoothly transmitted between any two vertexes in property graph-based data.

In order to ensure the correctness of the query, CORES has to record the bitset of all nodes along the query path and only enables the layer-by-layer bitset delivery. When traversing through each node again, the bitset is subjected to an $and$ operation with previously saved bitset. However, as the nested layers deepens and transmitting path lengthens, layer-by-layer delivery
will inevitably incur additional IO overhead and computation overhead. This is due to the engine's need to access corresponding arrays from external storage and perform bitset merging operations at each layer, which significantly decreases the efficiency. In the following, we introduce QUEST's novel skipping scheme based on the \textit{Skip-Tree} and basic bitset delivery operations to expedite scan-intensive query evaluating on unified nested structures by efficiently pruning the scan of most irrelevant instance.

\subsection{\textit{Skip-Tree} Based Payload Delivery}
In this section, we demonstrate the pivotal role of unified nested modeling with \textit{Skip-Tree} structure when evaluating the scan-intensive cross-model queries, showing how QUEST's innovative skipping scheme effectively reduces the I/O and memory cost. As for payload delivery on nested document-based data, we propose an efficient lowest common ancestor search algorithm based on the \textit{Skip-Tree} index. When it comes to pattern matching queries on graph data, the intricate mapping relations between attributes necessitates popular graph query languages like Cypher \cite{francis2018cypher} to impose limitations on traversal order. Hence, when furnished with the appropriate query path based on graph query languages like Cypher, we show how can QUEST's skipping scheme significantly reduce the query cost incurred by traversing the graph.

\subsubsection{\textit{Skip-Tree} based bitset transmitting from a node to a given ancestor}
Upon reviewing the fundamental bitset passing algorithm, the primary operation of transmitting intermediate results is to deliver the child's bitset to their parents. However, the layer-by-layer bitset transmitting results in additional I/O and computational costs. The proposed \textit{Skip-Tree} index can transmit the bitset of descendants to the ancestor directly by once bit calculation based on the pre-computed \textit{Skip-Counter}. As the \textit{Skip-Tree} operates like a \textit{Skip-List} on the query path from leaf node to its ancestor, the overhead of payload delivery is effectively reduced from $d$ to $\log d$ (where $d$ is the depth between leaf node to it's ancestor in hierarchy). 
Considering that the computation consumption is related to the cardinality of data instance in realistic use, which may make the efficiency promotion even more pronounced.

The skipping scheme based on \textit{Skip-Tree} is generalized by two pair-wise operations $\textit{SkipUp}$ and $\textit{SkipDown}$ which share the common essence with $\textit{RollUp}$ and $\textit{DrillDown}$. Here we take the \textit{Skip-Tree} structure shown in Figure \ref{skipcounter} based on the schema in document $S$ in Figure \ref{example} (a) as an example to illustrate the advanced bitset transmit operations $SkipUp$ and $SkipDown$ in our skipping scheme. Suppose we have a query defined in SQL-like form on $S$:

\noindent\textbf{SELECT} A.Email   \\
\textbf{FROM} Advertiser as A \\
\textbf{WHERE} A.Campaign.Wordset.Word = W \\
\textbf{AND} A.Campaign.Clicks.Person = P

We first begin the query processing from $\textit{Word}$. Suppose that $w_1$, $w_4$ satisfy the filtering condition in the $\textit{Word}$ instance, so we first get the bitset of $\textit{Word}$ as $10010000$. Then we transmit the bitset to $Campaign$ through the pre-calculated $\textit{SkipCounter}$ by $\textit{SkipUp}$ operation, where the bitset are merged into $110$. Next we transmit the bitset down to $Person$ through $SkipDown$ operation and get $1111000$ which means that we only need to scan the first four instances of the data and skip the judgement of last three instances. Suppose only $p_4$ satisfy the filtering condition in the first four $Person$ instances so we get the bitset $0001000$, and finally $\textit{SkipUp}$ to $Advertiser$ and fetch the corresponding $Email$ based on the bitset $10$ to get the correct output result $e_1$ of the example query.

\begin{figure}[h]
\centering
\includegraphics[height = 3.5cm]{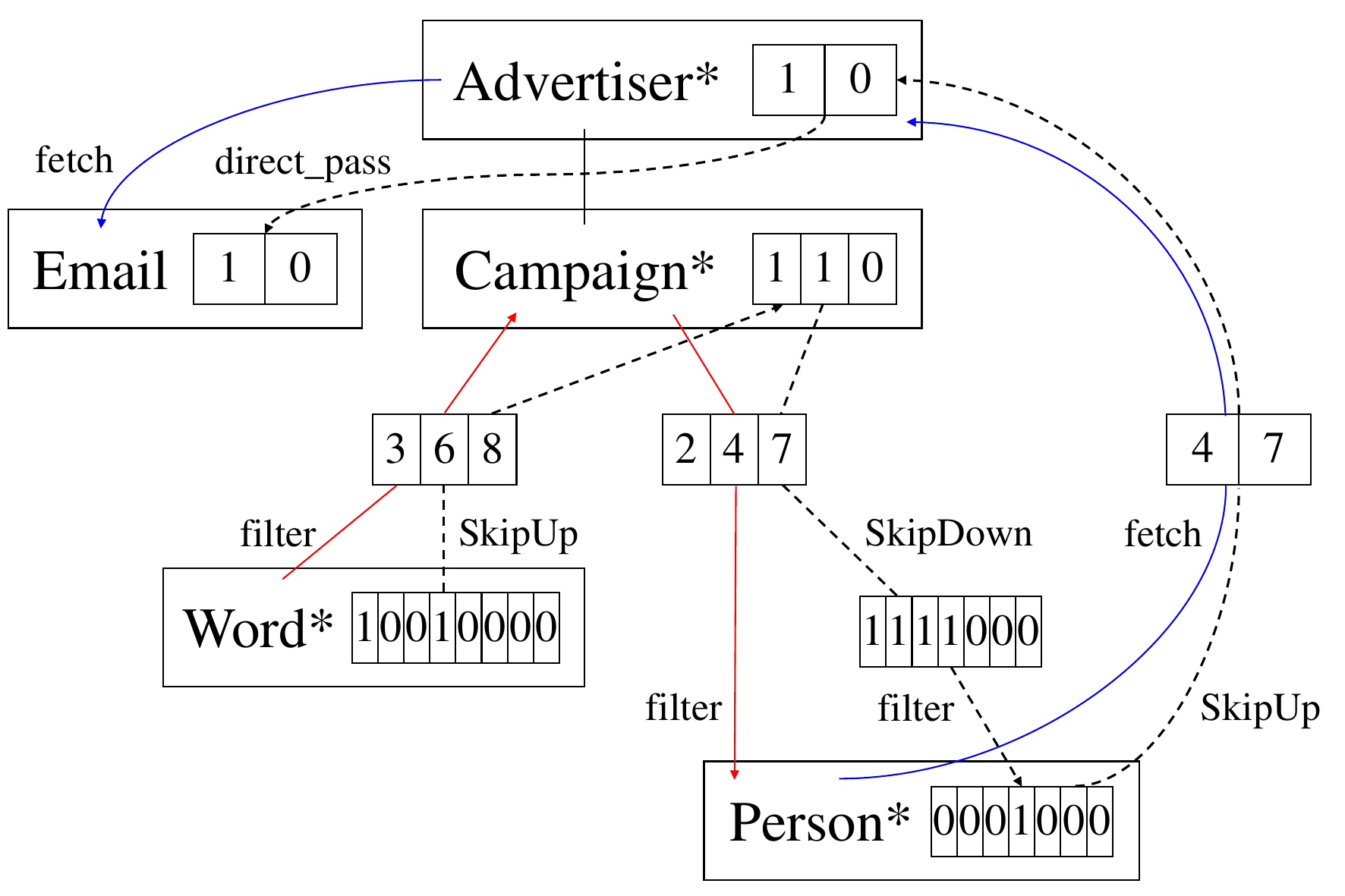}
 \caption{Illustration for SkipUp and SkipDown operation}
 \label{skipupanddown}
\end{figure}

The skipping scheme on the graph is similar to the nested document. The bitset delivery process based on \textit{Skip-Tree} still restores the pre-computed \textit{Skip-Counter} and \textit{Skip-Indicator} to the multi-hop connection matrix. By leveraging the aforementioned two-dimensional calculation akin to matrix multiplication, one can obtain a direct connection matrix that contains the intricate many-to-many mapping relationship between descendant and ancestor nodes. This enables seamless bitset transmission from descendants to ancestors with just a single bit operation. Therefore, when provided with the correct subgraph matching path using graph query languages such as Cypher, QUEST can efficiently transmit the payloads along the path based on \textit{Skip-Tree} structure and prune the scan of most irrelevant instance.

\subsubsection{Skip-Tree based efficient lowest common ancestor search algorithm on nested structure}

Furthermore, we demonstrate an additional application of the \textit{Skip-Tree} index: expediting the search for the lowest common ancestor (LCA) in tree-like nested structures. CORES \cite{wen2019cores} and DREMEL \cite{afrati2014storing} emphasized the pivotal role of LCA in transmitting intermediate query results within nested structures. Specifically, in our skipping scheme, it is imperative to transmit the bitset between two filters via their LCA. To accomplish this, we must first search for the LCA of the two filters based on the \textit{Skip-Tree} index. In addition, during the search process, QUEST will synchronously compute the \textit{Skip-Counter} of both filters up to their LCA which enables us to seamlessly transmit the bitset via their LCA through only once $Skipup$ and $Skipdown$ operation, thus reducing the extra I/O and computing costs caused by the layer-by-layer bitset transmitting.

\begin{algorithm}[]
\DontPrintSemicolon

  \KwInput{$n_s:$ Starting node; $n_d:$ End node; $ST:$ the \textit{Skip-Tree} Index.  }
  
  \KwOutput{$LCA:$ the lowest common ancestor of $n_s$ and $n_d$.}
\If{$n_s.depth>n_d.depth$}
{$v=n_s,\ s=n_d$}
\Else{$v = n_d,\ s=n_s$}

 \If{$s.isAncestor(v)$}{\KwRet \textit{$s$;} }
 $i = $ \textit{$ v.height$;} \;
 \While{$i\geq 0$}{
\If{$v.SkipAncestor[i].depth$$=$ \textit{$s.depth$}}
{$v$= $v.SkipAncestor[i]$;\;
\Break;
 }
\ElseIf{\textit{$v.SkipAncestor[i].depth
$}  $>$ \textit{$ s.depth$}}{\textit{$v$} = \textit{$v.SkipAncestor[i]$}; \;
 $i = v.height$; \;}
\Else{$i--;$}}
$j$ = \textit{$s.height$}; \;
\While{$j \geq 0$}{
\If{\textit{$v.SkipAncestor[j]$} $\neq$ \textit{$s.SkipAncestor[j]$}}
{\textit{$v$} $ =$ \textit{$v.SkipAncestor[j]$};\;
\textit{$s$} = \textit{$s.SkipAncestor[j]$}; \;
 $j$ = \textit{$v.height$}: \;}
\Else{$j--$;}

}
 
 \KwRet \textit{$v.SkipAncestor[0]$};

\caption{LCA searching algorithm based on Skip-Tree}
\label{alg:algorithm2}
\end{algorithm}

Algorithm \ref{alg:algorithm2} describes this process in detail. Given any two filters, 
we first compare the depth of the two filters in the nested structure to determine the deeper node, and let it skip to its ancestor node with the same depth as the shallower node based on its \textit{Skip-Ancestor} list. In the initialization of \textit{Skip-Tree} nodes' height, we employ a fixed lifting strategy which ensures the nodes at the same depth in the nested structure have equal heights within the \textit{Skip-Tree}. Therefore, nodes that are at the same depth in the nested structure can skip up synchronously based on their \textit{Skip-Ancestor} list. If two nodes do not reach the same ancestor after one step skip at high layer, we allow them to iteratively skip upward, otherwise we shall lower them by one layer. Repeat the above judgment until reach the very bottom in \textit{Skip-Tree}, thereby ensuring that the search result corresponds to the lowest common ancestor of the two filters.

Take the 4 layer \textit{Skip-Tree} structure in Figure \ref{skiptree} as an example, without loss of generality, assuming that the query payload is needed to transmit the bitset from node 14 to node 17. First, it is easy to determine node 14 is the deeper one. And the node 14's height in \textit{Skip-Tree} is 3, whose \textit{Skip-Ancestor} list is $\textit{[13,\ 12,\ 7,\ 0]}$, while the node 17's height is 2 with a SkipAncestor list$\textit{[16, 15, 0]}$. Then, by traversing down from the highest layer in the \textit{Skip-Ancestor} list of node 14 we find that node 7 is at the same depth as node 17 in the nested structure. Furthermore, they both appear highest at the $2^{nd}$ layer in \textit{Skip-Tree}, at where they will meet at the node 0 if they both take a step up. According to algorithm \ref{alg:algorithm2}, we shall lower them by one layer to the height of 1, where their \textit{Skip-Ancestor} are node 5 and node 15 respectively. Thus we take a step up at this layer and followed by lowering node 5 and node 15 to the bottom of the \textit{Skip-Tree} since they will again meet at the node 0 if they both take one more step up at layer 1. Therefore, we end the searching process at the node 4 at the bottom of \textit{Skip-Tree} which is exactly the LCA of node 14 and node 17. In the realistic use, QUEST will synchronously calculate the \textit{Skip-Counter} of node 14 and node 17 up to node 4 during the process of skipping upwards to search for the lowest common ancestor, thus the bitset of node 14 can be seamlessly transmitted to node 7 via their LCA node 4 by only once $SkipUp$ and $SkipDown$ operation, simultaneously reducing the extra I/O and computational cost caused by layer-by-layer delivery.

\subsection{Cross-Model Payloads Delivery}

In this section, we demonstrate QUEST's innovative scheme to enable seamless query payloads delivery on scan-intensive cross-model analytical workloads which significantly reduce the size of intermediate result. According to previous sections, QUEST establishes a unified logical representation based on the extended recursive definition of nested model, and develops a bitset-based skipping scheme based on \textit{Skip-Tree}. Therefore, the key to deliver the query payloads across different models is to efficiently transmit bitset across different nested structures through the join key.
This can be realized by constructing a new  $T_{indicator}$ to preserve the units matchup information between columns of the joinable attributes in different data models. In this way, the payload can be  seamlessly delivered  between two joined nested structure through the $T_{indicator}$ based on the bitset-based operations (\textit{i.e.}, $\textit{SkipUp}$ and $\textit{SkipDown}$), which enables QUEST's skipping scheme to efficiently evaluate the cross-model analytical queries.
In addition, before performing join algorithm (\textit{e.g.,} hash join in QUEST's implementation) to construct new $T_{indicator}$, QUEST will first transmit the bitset representing the previous payload to the join key and only apply the join algorithm(\textit{e.g.,} set up a hash table) to valid instances (\textit{i.e.,} the corresponding unit in bitset is set to 1), which further reduces the size of the intermediate results. For example, suppose there is a joint analysis on graph $G$ and document $S$ shown in Figure \ref{example}, where $G$ can be joined to $S$ through $PID$. Based on previous definition, $G$ can be modeled as nested tree structure as shown in Figure \ref{schema}, where \textit{Person}'s field type is \textit{record} since it's the root of the tree. If we construct a new $T_{indicator}$ based on the unit mathup information in the two columns of $Person.PID$ instances in $S$ and $G$, and embed it with the $T_{record}$ of \textit{Person} in $G$, the nested tree structure of $G$ shown in Figure \ref{schema} can be viewed as a subtree attaching to \textit{PID} in $S$, whose nested tree structure is shown in Figure \ref{example} (a). In this way, the query payload can be seamlessly delivered between $S$ and $G$ based on QUEST's skipping scheme.

\section{QUERY EXECUTION COST MODEL AND OPTIMIZATION}

\begin{table*}[h]
  \begin{center}
    \begin{tabular}{c c c c}
    \hline
    Parameter & Explanation & Parameter & Explanation\\
    \hline
    $T$ & Tree schema & $V$ & Node set on $T$\\ 

    $p_v$&Parent node of node&
    $D_v$ & Subtree interval of node $v$\\
    $G_v$ & Cardinality of node $v$ &  $B$ & Block size \\
    
    $F \subseteq V$ &  Filtering set&
    $P\subseteq V$ &  Fetching set \\
   
    $\sigma_v$& Selectivity of node $v$ in F&$\sigma_T= \prod_{v\in F}\sigma_v$&Total selectivity\\
    
    $O_F$& A sorted sequence of nodes in $F$& $W$ & Wandering sequence determined by $O_F$\\ 
    
    $\sigma_v(W)=\prod_{i=1}^{n_v}\sigma_i$ & cumulative selectivity until $v$ in $O_F$&  $\mathcal{W}$& Node set appeared in $W$\\ 
    
$S_v$&Average size per unit of node $v$& $m$& Average size per unit of \textit{Counter} or \textit{Indicator}\\
    
    $C_D(v)$ & Deserialization cost per unit on node $v$& $C_B(v)$ & Average bitset delivery cost per unit on node $v$\\

    $C_F(v)$& Average filtering cost per unit on node $ v $& $C_O(v)$ & Average regeneration cost per unit on node $v$\\

    $\#r_v(W)$&Runs of \textit{$RollUp$} on node $v$ in $W$&$\#d_v(W)$& Runs of \textit{$DrillDown$} on node $v$ in $W$\\
   \hline
    \end{tabular}
    \caption{List of the Involved Symbols and Parameters}
    \label{parameter}
  \end{center}
\end{table*}

\subsection{Query Evaluating Cost Modeling}
In this section, we first establish a comprehensive cost model of QUEST in evaluating analytical queries in nested tree structure under the situation of stripping the \textit{Skip-Tree} index since it has no impact on the setting up the optimization objective. Based on the unified nested tree model and columnar storage layout we established above, the scan-intensive cross-model queries evaluation can be logically regarded as a wandering through each predicate node in the expanded nested tree after performing the joins as described in section 3.3. During which, QUEST accesses the corresponding nodes following the order of wandering, scans each instance columns, performs the filtering operations, and delivers the intermediate result in the form of bitset along the wandering sequence. Since QUEST can efficiently prune the scan of most irrelevant instances based on payload delivery, the lower selectivity results in less I/O cost, an intuitive idea is to determine the order of predicate execution according to the level of selectivity \cite{wen2019cores}.
However, the wandering order may also greatly affect the overall efficiency of the query evaluating, since each wandering step in the nested tree during the bitset delivery requires reading the corresponding \textit{Counter} or \textit{Indicator} array from disk, which will incur additional I/O and CPU cost. Thus, when optimizing the predicates execution order of the scan-intensive cross-model analytical  workloads based on QUEST, it is also necessary to consider the length of the wandering path. Based on the notation in table \ref{parameter}, the I/O cost of query evaluating can be modeled as follows:
\begin{equation*}
\begin{split}
&C_{IO}=\sum_{v\in F}\frac{G_v\sigma_v(W)S_v}{B}+\sum_{v\in \mathcal W\setminus F}\frac{G_{p(v)}\#r_v(W)m}{B}+\sum_{v\in P}\frac{G_vS_v\sigma_T}{B}\\
\end{split}
\end{equation*}

The $C_{IO}$ mainly includes the cost of pruned scanning the columns of instance and metadata (\textit{i.e.,} \textit{Counter} and \textit{Indicator}) along the wandering sequence. Likewise, the CPU cost in the query plan can be modeled as follows:
\begin{equation*}
\begin{split}
C_{CPU}&=\sum_{v\in F}C_D(v)G_v\sigma_v(W)+\sum_{v\in \mathcal W\setminus F}
\left(G_{p(v)}\#r_{v}(W)+G_v\#d_v(W)\right)\\
& +\sum_{v\in F}C_F(v)G_v\sigma_v(W)+\sum_{v\in P}C_D(v)G_v\sigma_T+\sum_{v\in P}C_O(v)G_v\sigma_T\\
\end{split}
\end{equation*}

The $C_{CPU}$ mainly includes the cost of  predicates evaluating, bitset computation, and records deserialization and regeneration along the wandering sequence. Thus, the total query evaluating cost can be modeled as: $Cost(W)=C_{IO}+C_{CPU}$. Thus, optimizing the total query cost is equivalent to solve for $W$, such that: $W = \arg\min_{W}Cost(W)$.

\subsection{Correctness Constraint For Predicate Execution Ordering}
However, $W$ can't be arbitrary since delivering the payload in an incorrect predicate order may lead to filtering information loss and result in incorrect query results. This is because if we \textit{$RollUp$} from a node to his parent and then \textit{$Drilldown$} back to it, we can't get the original bitset in most of the cases since there is a one-to-many mapping relation. One way to solve the problem is to record the bitset of all nodes along the query path and when traversing through each node again, the bitset is subjected to an $and$ operation with previously saved bitset \cite{wen2019cores}. However, this extra overhead can be erased by constraining the wandering sequence $W$ which also enables the query evaluation to leverage the power of QUEST's novel skipping scheme based on \textit{Skip-Tree} to further reduce the I/O and CPU cost\footnote{There still exists the case that requires QUEST to record the bitset in a small number of nodes during the wandering in graph-based nested structure, we omit the specific analysis since it has no affect in applying the skipping scheme, and costs little.}.
Formally, the wandering sequence should satisfy the following constraint:
\begin{equation*}
\begin{split}
W:&\ w_1w_2...w_n\ is\ a\ wandering\  sequence\ on\ T,\ uniquely\\
&determinted\ by\ O_F\ (a\ sort\ sequence \ of\ filters\ in\ F),\\
&restrained\ by: \forall k\in[i,j],w_k\in D_v\ when\ w_i=w_j=v.
\end{split}
\end{equation*}

An intuitional interpretation of the constraint is: if the bitset is transmitted out of one node's subtree, the payload delivery process should never back to its subtree again. That is, when transmitting bitset through a node, it should be ensured that all of the filtering predicates in its subtree must already have been evaluated. In this way, the bitset delivery process between any two filtering predicates is ensured to exactly via their most lowest common ancestor. In next section, we will consider the predicate ordering optimization method under the above constraints and explore the efficient heuristic algorithm.

\subsection{A Heuristic Algorithm For Query Execution Plans Optimization}
To explore efficient algorithm for solving optimal predicate ordering problem, we first simplify the above cost model as:
\begin{equation*}
\begin{split}
 Cost(W)& = A \sum_{v\in F}G_v\sigma_v(W)+ \\ & \quad\quad B\sum_{v\in \mathcal W\setminus F} \left(G_{p(v)}\#r_{v}(W) + G_v\#d_v(W)\right),
\end{split}
\end{equation*}
where $A,B$ are constants uncorrelated to $W$. And the query optimization problem can be defined formally as follows:
\begin{flalign}
&\nonumber \textbf{Input:}\ A, B, G_v, \sigma_v, p(v), \#r_v, \#d_v, F, \text{correctness constraint}\ I  && \\
&\nonumber \textbf{Output:}\ W=\arg\min_W Cost(W),\ \text{where}\ W\ \text{satisfies}\ I. &&
\end{flalign}
Based on the simplified problem, we introduce a feasible heuristic algorithm shown in algorithm \ref{alg:algorithm3}, where 
the wandering process can be viewed as a selectivity-aware postorder traverse. This is because the selectivity of filtering predicates takes a important role to reduce the query evaluation cost according to QUEST's bitset-based query pushdown and payload delivery. Meanwhile, according to previous correctness constraint of $W$, we shall complete the evaluation on all of the filtering predicates in a node's subtree interval before we wander out of it. Therefore, as illustrated in algorithm \ref{alg:algorithm3}, we begin the wandering from the most selective filtering predicate and back to root step by step. During the process, if there exist filtering predicates haven't be evaluated yet in one node's subtree interval, we again choose the most selective one to achieve first and continue the wandering to back to root step by step from it. By iteratively execute the searching process until the wandering get back to root and traversed all of the filtering predicates, we could obtain an $O_F$ that satisfy the correctness constraint $I$. Thus, we could directly perform the \textit{Skip-Tree} based LCA searching algorithm as well as the $\textit{SkipUP}$ and $\textit{SkipDown}$ functions to correctly deliver the payload and output the result. While it is important to analyze the complexity of this problem and explore an optimal query execution plan, we leave this to future work due to space and time constraints.

\begin{algorithm}[]
\DontPrintSemicolon

  \KwInput{$T$: Tree schema; $V$: Node set on T; $F\in V$: Filtering set; $\sigma _v$: Selectivity of node $v$ in $F$; $D_v$: Subtree interval of node $v$;}
  
  \KwOutput{$O_F$: a sort sequence of nodes in F}

 $O_F = []$, $s_r$ = the root of $T$;\;
 $s$ = the most selective node in $F$;\;
 $O_F.\textit{append}(s)$;\;
 \While{$s \neq s_r$ or $\textit{IsFiltered}(s)==\textit{Nil}$}{
 \If{$IsFiltered(D_{s})$}{
 $s = s.Father$;\;
 
 }
 \Else{$s = FindMSelective(D_{s})$;\;
 $O_F.append(s)$;}
 \tcp{Find the most selective filter that haven't been traversed in $D_{s}$}
 }

\KwRet $O_F$
  
\caption{Heuristic algorithm for optimal filtering order}
\label{alg:algorithm3}
\end{algorithm}

\section{EXPERIMENTAL EVALUATION}

In this section, we demonstrate  QUEST's high efficiency of evaluating scan-intensive cross-model analysis through detailed experiments.
We utilize a single-node server which has a AMD Ryzen 9 5900X 12-Core CPU, 1TB WD Black SN750/PC SN730 NVMe SSD disk with about 3000MB/s and 500MB/s performance for sequential and random read and 2 $\times$ 32g VENGEANCE LPX ddr4 3200hz memory.

\subsection{Dataset and Workloads}It is non-trivial to design the workloads which not only cover the most important paradigms of cross-model query processing but also simulate realistic use cases \cite{zhang2019unibench}. However, we believe that the data schema in UniBench \cite{zhang2019unibench} is not complex enough to reflect the structural characteristics of multi-model data in realistic use, especially in nested documents and property graphs, which will limit the comprehensiveness and depth of experimental evaluation. Thus, we put a lot of effort into generating more profound multi-model data sets for scan-intensive cross-model analysis according to the data schema depicted in Figure \ref{example}. Specifically, nested document-based data and relational table are generated to match the scale of standard LDBC data. The SF 30 LDBC graph dataset is matched with a 10 GB nested document-based data and a corresponding relational table size in 6 megabytes which has same entries scale as the person node in LDBC. Due to computational resource constraints, we only imported data relevant to the queries with a total size of 20.24 GB.

Based on newly generated multi-model data, we simulate a holistic analysis on users’
social behaviors and advertisers’ promotional campaigns. And referring to the choke points design of popular benchmarks such as LDBC \cite{angles2020ldbc} and TPC-H \cite{boncz2013tpc,dreseler2020quantifying}, we further refine the choke points of scan-intensive cross-model analysis.

\begin{figure}[h]
\centering
\includegraphics[height = 2cm]{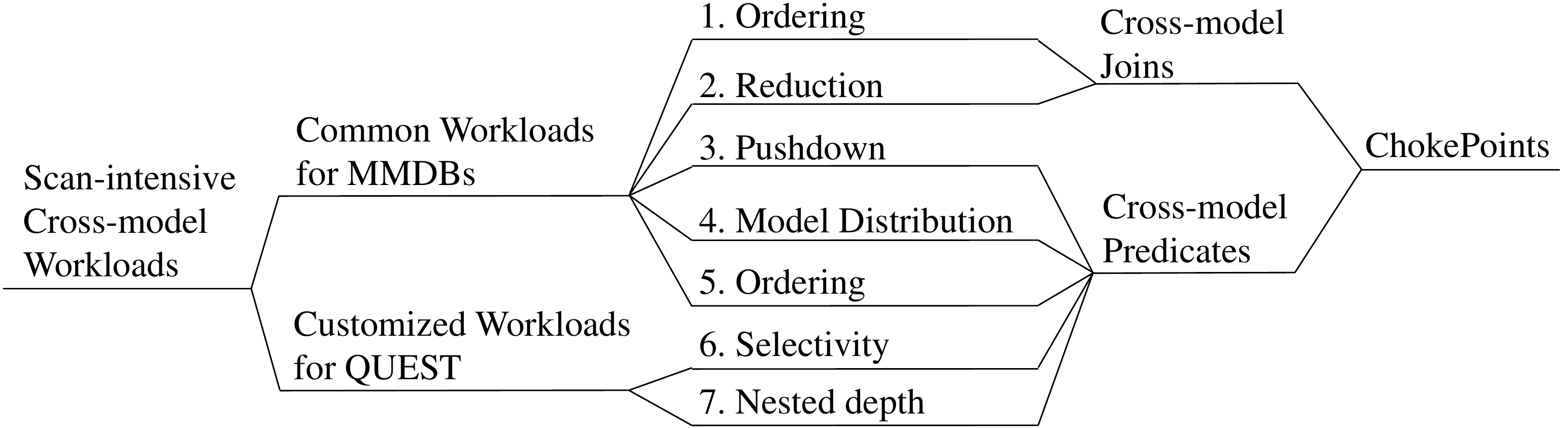}
 \caption{Choke points design }
 \label{chokepoint}
\end{figure}

Specifically, we group the choke points of scan-intensive cross-model analysis into two categories: 
First, \textit{cross-model joins related choke points} affect the efficiency of joining across different data types which includes join ordering and reduction. The scale reduction of intermediate result achieved by these optimizations is crucial in evaluating cross-model analysis. 
Second, \textit{cross-model predicates related choke points} affect the query evaluating efficiency in individual type of data. Among them, predicate pushdown and ordering are two highly influential key points at logical dimension in the TPC-H benchmark \cite{dreseler2020quantifying}, which are also the two key problems QUEST mainly focus on when evaluating scan-intensive cross-model analytical queries. The distribution of predicates in different models is also a key point affecting the processing efficiency of analytical queries in multi-model databases, since a single engine multi-model database can hardly achieve the state-of-the-art efficiency of all tasks in all models at the same time and there are always some trade-offs among different data types. According to the above five choke points, we design the corresponding scan-intensive cross-model queries for common MMDBs. However, in order to further test the efficiency of our proposed  evaluation scheme, comprehensively test the pros and cons of QUEST, we design customized query loads, focusing on the impact of selectivity and nested depth of predicates in nested structure on the query evaluating efficiency. Overall, the workload description is shown in table \ref{Workload} where the total selectivity of most queries is set to $5\%$ as high selectivity and  $10\%$ as low selectivity, while the deep and shallow nested depth means that the depth of the deepest predicate in the query is set to 7 and 3, respectly.    

The mainstream MMDBs are selected as the main comparison objects, including Arango-DB-community-3.10.8 \cite{arangodb} and OrientDB-community-3.2.20 
 \cite{orientdb} which are the two most commonly tested databases when it comes to cross-model workloads \cite{zhang2019unibench,van2023comparative}. In addition, we map multi-model data to relational paradigm and graph pattern by trivial data modeling, and conduct more detailed experiments on a column-orient relational database ClickHouse-community-23.5.3.24 \cite{ClickHouse} together with a graph database Neo4j-community-5.9.0 \cite{neo4j} which both show excellent analytical performance on individual data type, to further explore the pros and cons of various queries evaluation schemes in dealing with scan-intensive cross-model analysis. In all experiment, we use default indexes which are built on primary keys, and no secondary index is created. 

\begin{table}[h]
  \begin{center}
    \begin{tabular}{|c|c|c|c|c|}
    \hline
    Queries & R/D/G & Selectivity & Nested depth & Choke Points \\
    \hline
    Q1&2/2/2&high&deep&\textbf{1, 2, 3, 4, 5, 6, 7}\\
    \hline
    Q2&3/1/1&high&deep&1, 2, 3, \textbf{4}, 5, 6, 7\\
    \hline
    Q3&1/3/1&high&deep&1, 2, 3, \textbf{4}, 5, 6, 7\\
    \hline
    Q4&1/1/3&high&deep&1, 2, 3, \textbf{4}, 5, 6, 7\\
    \hline
    Q5&2/2/0&high&deep& \textbf{2}, 3, \textbf{4}, 5, 6, 7\\
    \hline
    Q6&2/0/2&high&deep& \textbf{2}, 3, \textbf{4}, 5, 6, 7\\
    \hline
    Q7&0/2/2&high&deep& \textbf{2}, 3, \textbf{4}, 5, 6, 7\\
    \hline
    Q8&2/2/2&low&deep& 1, 2, 3, 4, 5, \textbf{6}, 7\\
    \hline
    Q9&2/2/2&high&shallow& 1, 2, 3, 4, 5, 6, \textbf{7}\\
    \hline
    Q10&0/3/0&high&deep&  \textbf{3}, \textbf{5}, 6, 7\\
    \hline
    Q11&0/0/3&high&deep&  \textbf{3}, \textbf{5}, 6, 7\\
   \hline
   
    \end{tabular}
    \caption{Workload description (R/D/G represents the number of predicates in Relation/Document/Graph model)}
    \label{Workload}
  \end{center}
\end{table}

\subsection{Performance Evaluation}
\subsubsection{Query running time}

\begin{figure}[h]
\centering
\includegraphics[width=1\linewidth,height = 4cm]{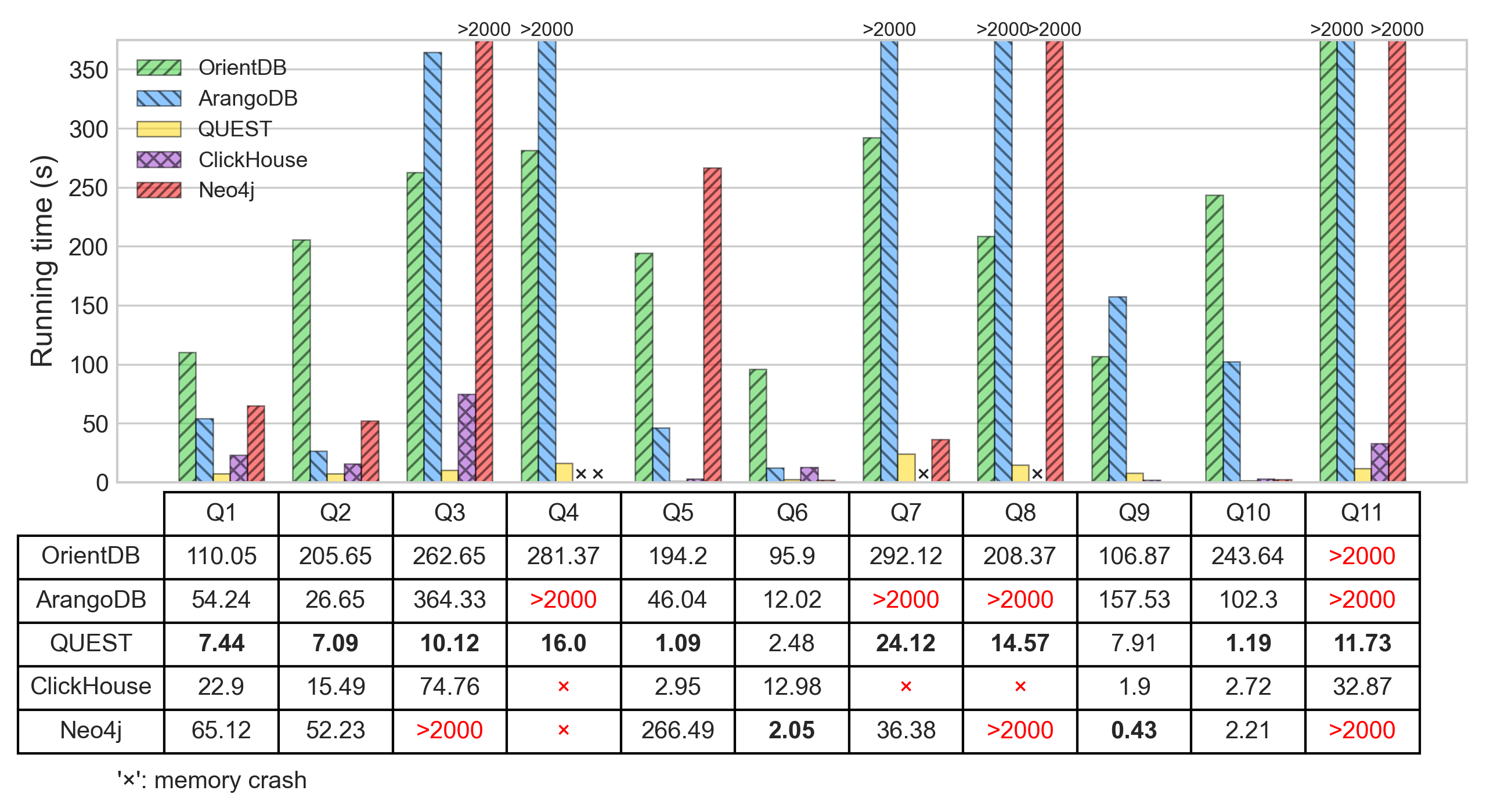}
 \caption{Query running time}
 \label{time}
\end{figure}
In all scan-intensive cross-model query evaluation experiments, QUEST out performs all MMDBs in query running time and improves the performance by $3.7 \times - 178.2\times$. It is only slightly slower than the two analytical databases Neo4j and Clickhouse in the query of $Q6 $ and $Q9$. QUEST has the most stable performance among all competitors as it keeps the running time under 25 seconds in all queries,
while the rest databases may incur severe query latency timeouts when evaluating specific scan-intensive cross-model queries (we set 2000 second as the upper time limit). The excellent analytical performance mainly gains from the column-orient skipping scheme which enables QUEST to push scan-intensive cross-model queries down to storage and prune the scan of most irrelevant instance.

\subsubsection{Memory usage}

\begin{figure}[h]
\centering
\includegraphics[width=1\linewidth,height = 4cm] 
{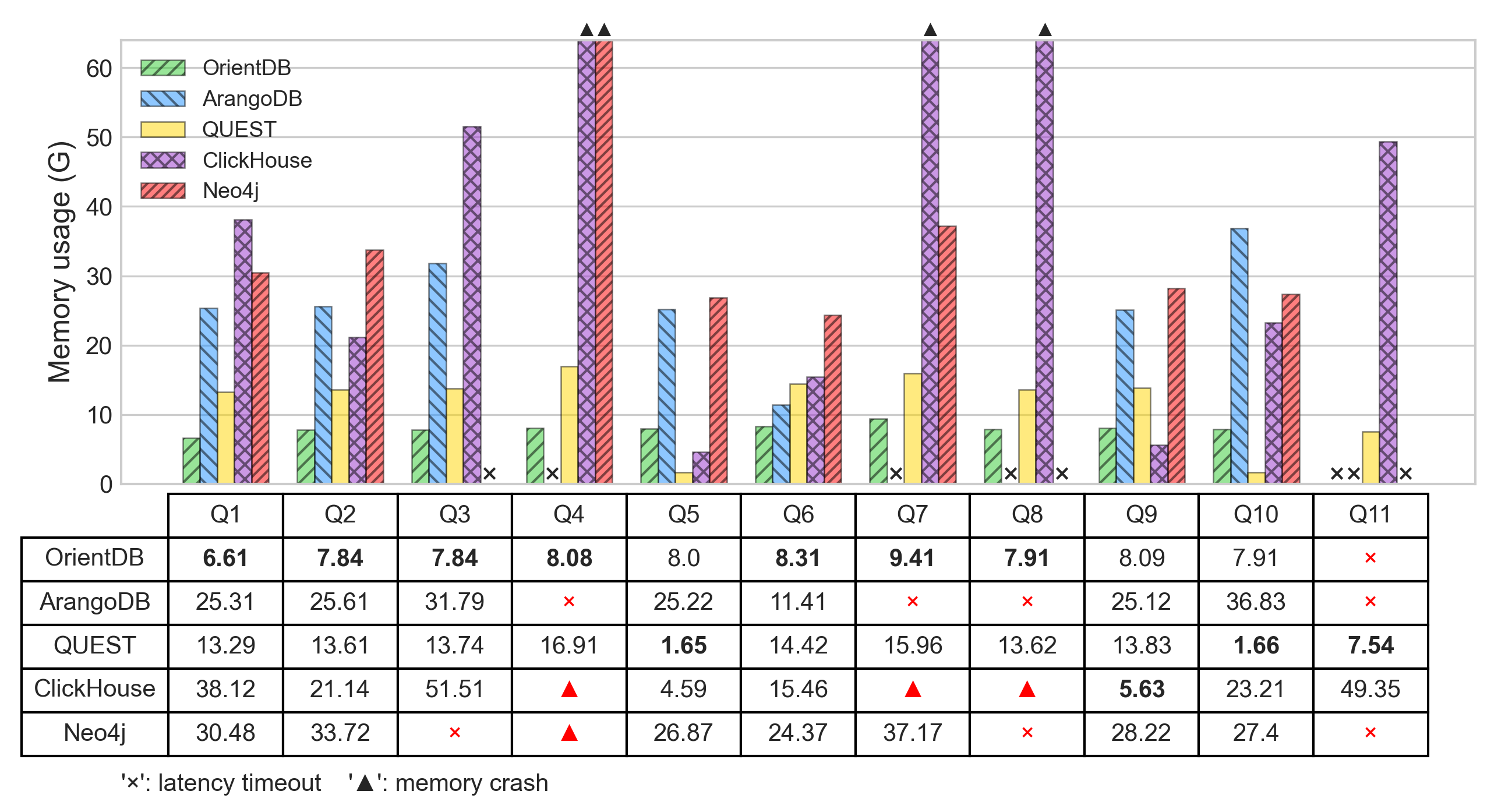}
 \caption{Memory Usage}
 \label{memory}
\end{figure}
Although the memory usage of QUEST is not the least in most of scan-intensive cross-model queries, it is  kept in a pretty low section between 1.66 GB to 16.91 GB, second only to OrientDB. In addition, QUEST is not like other competitors to generate huge intermediate results and incur memory crash in specific scan-intensive cross-model queries. 
This mainly credits to bitset-based payload delivery strategy, which significantly reduces the scale intermediate results.
The results verify QUEST's robustness in memory usage when evaluating the scan-intensive cross-model analysis. 
\subsubsection{Disk space overhead}

QUEST takes 17.83 GB disk space  to store 20.24 GB multi-model data which includes extra 2.85 GB space of \textit{Skip-tree} index. However, the disk space overhead  is still significantly lower than other MMDBs, second only to ClickHouse.
Although QUEST does not focus on applying advanced column-orient compression scheme to multi-model data, it naturally comes as an additional benefit of employ the unified columnar storage. We leave the efficient direct processing on compressed data as future work.
\begin{table}[h]
  \begin{center}
    \label{sidsk}
    \begin{tabular}{|c|c|}
    \hline
      Systems & Disk Space Overhead\\
    \hline
    OrientDB&44.56 GB\\
    \hline
    AgrangoDB&28.23 GB\\
    \hline
     QUEST& 17.83 GB\\
    \hline
    ClickHouse&6.91 GB\\
    \hline
    Neo4j& 36.22 GB\\
    \hline
    \end{tabular}
    \caption{Disk space overhead}
  \end{center}
\end{table}
\subsection{Detailed Analysis}

We next conduct a further profound analysis on the experimental results in conjunction with choke points to comprehensively evaluate the pros and cons of QUEST when evaluating scan-intensive cross-model queries.

\subsubsection{Predicates' model distribution.} The single engine MMDBs face the "one size doesn't fit all" dilemma when evaluating scan-intensive cross-model analysis. For example, Arango DB and OrientDB out perform Neo4j in the cross-model queries that mainly focus on document like $Q3$ and $Q5$, due to the underlying document-first citizen storage data layout. However, they are much slower than Neo4j when evaluating the graph-focus cross-model queries like $Q6$ and $Q7$. And all of them face latency timeouts or memory crash in more complex graph-focus cross-model queries like $Q4$ and $Q8$. Clickhouse remains excellent performance in scan-intensive cross-model queries of different predicates distribution, as it flattens the nested relation and storage in separate tables. But it meet with the most serious memory crash due to the most frequently joining. QUEST is most non-sensitive to the distribution of predicates. Particularly, although QUEST's unified representation is based on the extended nested structure, we find out that QUEST remain its high efficiency in evaluating the complex multi-hop query $Q11$ in graph, even out performs the specific graph database. This is mainly credits to the efficient pruning scheme based on the extended recursive definition of nested tree structure and the bitset-based payload delivery. 
Overall, the unified columnar layout enables QUEST to seamlessly deliver query payloads across different types of data and thus maintain excellent performance in scan-intensive cross-model queries with predicates in various model distribution. 
\subsubsection{Nested depth.}
The nested depth of predicate in QUEST's unified nested logical model greatly affects the efficiency of evaluating scan-intensive  cross-model queries. The deeper nested depth of predicates results in longer traversal path in graph-based data, more joins on relational tables and more complicated data retrieval in nested document-based data. Thus, we find out that most of the system has much better performance on $Q9$ than on $Q1$ where the total selectivity of two queries are controlled as the same. However, QUEST's performance even improves as the nested depth getting deeper. This mainly credits to the I/O reduction gains from \textit{Skip-Tree} index structure and the pair-wise operations $\textit{SkipUp}$ and $\textit{SkipDown}$. We conduct more specific ablation experiment to evaluate the benefits of \textit{Skip-Tree} structure in nested document-based data and graph-based data. The result shows that when we strip the \textit{Skip-Tree} from QUEST to evaluate $Q10$ and $Q11$, the query running time increase $17\%$ and $16\%$ as well as the memory usage increase $5\%$ and $2\%$. In addition, the nested depth in our data is too shallow to fully leverage the strength of \textit{Skip-Tree}. It's obvious that as the nested depth gets deeper, the absolute I/O reduction gains from \textit{Skip-Tree} will become even more prominent. Overall, the \textit{Skip-tree} enhances the robustness of QUEST when evaluating scan-intensive cross-model queries with deeper nested depth, in terms of both query execution time and memory usage.

\subsubsection{Selectivity}
The selectivity of queries has great affect on the efficiency of evaluating scan-intensive cross-model queries. As the selectivity increase, the size of intermediate result will significantly increase if the query evaluation scheme fall short to further reduce the unnecessary scan of data instance. Comparing the query running time of Q1 and Q8, we observe that as the selectivity of the queries increases, most systems suffer sever query response latency or memory crash. Although the query running time of QUEST also increases from 7.44s to 14.57s and the memory usage increases from 13.29 GB to 13.62 GB, they are still be kept in a very low level. This mainly credits to the bitset-based payload delivery, as it greatly reduce the size of intermediate result and enables efficiently pruning on evaluation of scan-intensive cross-model queries. Thus, the bitset-based payload delivery enhances the
robustness of QUEST when evaluating scan-intensive cross-model
queries with higher selectivity, in terms of both query evaluating time and memory usage.

\subsubsection{Summary} Detailed experiments verify the efficiency of QUEST in evaluating scan-intensive cross-model queries. This is reflected in less response time, lower memory usage, and lower disk space overhead. QUEST also shows excellent robustness when faced with predicates in various model distributions, regardless of their nested depth and selectivity. 

\section{RELATED WORKS}
Extensive studies have been conducted to develop systems for both cross-model and scan-intensive queries evaluating in industry and academia which evolves into multi-model databases and column-orient databases nowadays. However, after reviewing the two trend, we demonstrate that most of their query evaluation schemes fall short to efficiently process scan-intensive cross-model analysis.

\textbf{Cross-model queries evaluation.}
The variety of data is one of the most challenging issues for the research and practice in data management systems \cite{lu2019multi}. A number of systems are developed to manage multi-model data and thus evaluate cross-model queries. One of them is the
middleware-based multi-engine database, which can mainly be further divided into two categories: 
(1) Multistore system. It contains multiple heterogeneous data stores and a unified query interface. Representative systems are HadoopDB \cite{abouzeid2009hadoopdb}, Estocada \cite{bugiotti2015invisible, alotaibi2019towards} and Polybase \cite{dewitt2013split}. Forresi et al. proposes a framework to support data analysis within a high-variety multistore  through a dataspace layer \cite{forresi2021dataspace}. (2) Polystore system. It contains multiple heterogeneous data stores and multiple query interfaces \cite{tan2017enabling}. DBMS+ \cite{lim2013fit}, for example, embraces multiple processing and database platforms through unified declarative processing. BigDAWG \cite{duggan2015bigdawg} presents a polystore architecture based on the island of information design. Other representative systems are RHEEM \cite{agrawal2018rheem}, and Myria  \cite{wang2017myria}. However, the efficiency of these systems' scan-intensive cross-model queries evaluation schemes are greatly limited, due to the need of joinning different types of data stored in  distinct places, which results in extra
data copy, migration and integration costs.  

The systems based on single-engine optimize the query processing performance by establishing a fully integrated backend \cite{lu2018udbms}. Depending on the underlying storage engine, they can be further classified as follows: 
(1) Relation-based storage. Many relational databases have extended underlying compatibility and optimization to other data types, such as PostgreSQL \cite{postgresql}, DB2 and Sinew \cite{tahara2014sinew}. (2) Document-based storage. ArangoDB \cite{arangodb} and MongoDB \cite{mongodb} are multi-model databases that support key-value, document and graph data. (3) Graph-based storage. OrientDB \cite{orientdb} is an open source NoSQL DBMS that supports graph, key-value, document, and object models. Although these systems achieve unified storage for multi-model data, their queries evaluation schemes are inappropriate
for efficiently processing scan-intensive cross-model analysis, as the giant intermediate result may result in response latency and process crash. There is still a lack of customized unified data layout and query evaluation scheme for can-intensive cross-model analysis. 

\textbf{Scan-intensive queries evaluation.}
Column-orient databases has been developed to reduce the IO and CPU cost of typical analytical queries. Due to its excellent performance on processing scan-intensive queries, it has been extended to multiple data models to speed up the processing of analytical queries. Depending on the type of stored data, columnar storage systems can be classified as follows: 
(1) Columnar storage system for relational tables. Specific optimizations include virtual row IDs \cite{nes2012monetdb}, block-oriented and vectorized processing \cite{boncz2005monetdb, abadi2006integrating}, columnar compression and direct manipulation of compressed data \cite{abadi2006integrating, abadi2006materialization, zukowski2006super}, efficient join operations \cite{abadi2006integrating, manegold2004cache}. (2) Column query systems for documents. DREMEL \cite{melnik2010dremel} first introduced the column storage management approach to nested document data \cite{afrati2014storing}, and Steed \cite{wang2017exploiting} has since optimized simple paths based on it. CORES \cite{wen2019cores} designed a new regeneration embedding scheme to deliver query payload along arbitrary path in nested schema and further reduce query overhead. AMAX \cite{alkowaileet2022columnar} optimized DREMEL's embedding scheme and leverage LSM structure to support efficiently data updates \cite{alkowaileet2022columnar}. (3) Column storage system for graph data. Existing graph databases usually store graph topology in columnar structure \cite{gupta2021columnar}. Gupta et al. presented a pure columnar storage and list-based processing for main memory GDMS \cite{gupta2021columnar}. There are also some works to extend columnar relational databases to accelerate the analytical queries on graph, such as GRainDB \cite{jin2021making}, GQ-Fast \cite{lin2016fast}, \textit{etc.} (4) Columnar storage compatible with multi-model data. Only a few columnar storage databases are currently compatible with both relational tables and document data, such as CrateDB \cite{cratedb01} and HPE Vertic \cite{hpevertic}. SAP HANA \cite{lee2017parallel, sherkat2019native} is a HTAP-friendly commercial database that supports multi-model columnar storage, but it is an in-memory database and not open source. Although the efficiency of column-orient query evaluation schemes have been proven in various types of data, it's challenge for them to continue their excellent performance on cross-model analysis. QUEST aims to tackle the challenges and present an efficient query evaluation scheme towards scan-intensive cross-model analysis.

\section{CONCLUSION AND FUTURE WORK}
This paper presents QUEST which is an efficient
query evaluation scheme towards scan-intensive cross-model analysis. QUEST employs columnar layout to unify the representation of relational, nested document based and property graph-based data. 
The detailed experiments show that QUEST has excellent and stable performance in evaluating scan-intensive cros-model queries, which is reflected in less response time, lower memory usage and reduced disk space overhead when faced with predicates in various model distribution, regardless of their nested depth and selectivity.

As for future work, we shall incorporate more data models into QUEST, including but not limited to spatial data, time series and vector, \textit{etc}. Next we shall extend QUEST to be HTAP-friendly based on log structure to support efficient multi-model data modification and ACID properties of transactions. Futhermore, we plan to pursue new cross-model query optimization opportunities to make QUEST even faster and also scale to distributed execution.

\newpage

\bibliographystyle{ACM-Reference-Format}
\bibliography{sample-base}

\end{document}